\documentclass{emulateapj}

\newcommand{\beq}{\begin{equation}}
\newcommand{\eeq}{\end{equation}}

\def\arcmin{\hbox{$^\prime$}}
\def\arcsec{\hbox{$^{\prime\prime}$}}
\def\solar{\mbox{$_{\normalsize\odot}$}}

\newcommand{\AmS}{{\protect\the\textfont2
  A\kern-.1667em\lower.5ex\hbox{M}\kern-.125emS}}
\newcommand{\lsim}{\ \raise
-2.truept\hbox{\rlap{\hbox{$\sim$}}\raise5.truept\hbox{$<$}\ }}
\newcommand{\gsim}{\ \raise
-2.truept\hbox{\rlap{\hbox{$\sim$}}\raise5.truept\hbox{$>$}\ }}
\newcommand{\simsim}{\ \raise
-2.truept\hbox{\rlap{\hbox{$\sim$}}\raise5.truept\hbox{$\sim$}\ }}

\slugcomment{Accepted for Publication to ApJ}

\shorttitle{The IMF of LH~95 in the LMC down to the Sub-Solar Regime}

\shortauthors{Da Rio, Gouliermis, \& Henning}

\begin{document}

\title{The Complete Initial Mass Function Down to the Sub-Solar Regime
in the Large Magellanic Cloud with {\sl Hubble Space Telescope} ACS
Observations\altaffilmark{1,}\altaffilmark{2}}

\author{Nicola Da Rio\altaffilmark{3}, Dimitrios A. Gouliermis, and
Thomas Henning}

\affil{Max-Planck-Institut f\"{u}r Astronomie, K\"{o}nigstuhl
17, D-69117 Heidelberg, Germany;\\ dario@mpia.de, dgoulier@mpia.de,
henning@mpia.de}


\altaffiltext{1}{Based on observations made with the NASA/ESA {\em
Hubble Space Telescope}, obtained at the Space Telescope Science
Institute, which is operated by the Association of Universities for
Research in Astronomy, Inc. under NASA contract NAS 5-26555.}

\altaffiltext{2}{Research supported by the German Research
Foundation (Deutsche Forschungsgemeinschaft) and the German Aerospace
Center (Deutsche Zentrum f\"{u}r Luft und Raumfahrt).}

\altaffiltext{3}{Member of IMPRS for Astronomy \& Cosmic Physics
       at the University of Heidelberg, Germany}


\begin{abstract}

In this photometric study of the stellar association
LH~95 in the Large Magellanic Cloud (LMC) we focus on the pre-main
Sequence (PMS) population in order to construct, for the first time,
the sub-solar initial mass function (IMF) in the LMC. The basis for this investigation
consists of the deepest photometry ever performed in the LMC with the {\sl
Advanced Camera for Surveys} (ACS) on-board the {\sl Hubble Space
Telescope} ({\sl HST}). We improve our catalog of point sources obtained
with ACS (Paper~I). We carry out
a Monte Carlo technique to subtract the contribution of the general field
of LMC. We isolate the
central region in the observed area of the association, as the part
characterized by the highest concentration of PMS stars. We analyze the
reddening distribution of the system in order to obtain the mean value of
extinction and we study the mass function of its field-subtracted
population. For this purpose, we introduce a new set of
evolutionary models, derived from the calculations on the evolution of PMS
stars by Siess and collaborators. We use these models with our
observations of LH~95 to derive the IMF of the system.  This mass function
is reliably constructed for stars with masses down to $\simeq
0.43$~M$_{\odot}$, the lowest mass ever observed within reasonable
completeness in the Magellanic Clouds. Consequently, its construction
offers an outstanding improvement in our understanding of the low-mass
star formation in the LMC. The system IMF of LH~95 shows a definite change in its
slope for masses $M\lesssim1$~M$_{\odot}$, where it becomes more shallow.
In general, the shape of this IMF agrees very well with a multiple
power-law, as the typical Galactic IMF, down to the sub-solar regime.
The change in the slope (``the knee'') of our IMF at $\sim 1$~M{\solar} also agrees with the average Galactic IMF.
As far as the
slope of this system IMF is concerned, it is found to be somewhat more shallow than the corresponding ``classical'' Galactic IMF in the sub-solar regime, probably due to unresolved binarity, while for stars with $M\gsim1M\solar$ it becomes slightly steeper.
We do not find significant differences in the shape of the overall
IMF of LH~95 from that of each of the three individual sub-clusters of the
association. This clearly suggests that the IMF of LH~95 is not subject to
local variability.
\end{abstract}


\section{Introduction}

The Magellanic Clouds, the closest undisrupted dwarf galaxies to
our own galaxy, host numerous young stellar associations. The
study of the low-mass populations of these systems provide important
improvements in understanding extragalactic star formation. The
investigation of the star formation processes, and the Initial Mass
Function (IMF), together with the time dependency of the formation events, are
key points in the characterization of stellar populations in any kind of
concentrations, from small clusters to entire galaxies.

Regarding the IMF, several studies have been carried out in the past for
the Galaxy \citep[e.g.][]{salpeter, scalo, kroupa, chabrier2003},
describing the initial numbers of massive stars
($M\gtrsim8$~M$_{\odot}$), stars of intermediate mass
($1$~M$_{\odot}\lesssim M \lesssim 8$~M$_{\odot}$), low-mass stars
($0.08$~M$_{\odot}\lesssim M \lesssim 1$~M$_{\odot}$) and brown dwarfs
($M\lesssim 0.08$~M$_{\odot}$). The findings of such investigations
support the hypothesis of the universality of the IMF, in the sense that
the average IMF measured in a diversity of galactic systems does not
show noticeable variations \citep{kroupascience}. This statement turns
out to be rather certain for the intermediate- and high-mass stars,
while for low-mass stars there could be a dependence of the IMF slope on
metallicity, in the sense that metal-rich environments tend to produce
more low-mass stars than metal-poor systems
\citep{piotto1999,reyle2001}, although several IMF measurements are
needed to confirm such a systematic effect. In low-mass studies there
are several uncertainties present. For example the study of the IMF in
metal-poor galactic globular clusters \citep{piotto1999} is
significantly limited by the corrections for dynamical evolution
effects. Specifically, phenomena such as mass segregation and unknown
binarity fraction change dramatically the presently measured mass
distribution with respect to the one at the time of the formation of
these systems \citep{marks08}.

Under these circumstances, ideal environments for the search
of a metallicity dependence of the IMF are the Magellanic
Clouds. Their stellar populations are characterized by
metallicities of $[Fe/H] \simeq 2.5$ to $5$ times lower than in the disk
of the Milky Way \citep{luck99}, but have a similar star formation rate
(SFR) \citep{westerlund97}. Furthermore, the low metal abundances of the
Magellanic Clouds suggest that their environments $-$ as well as
in other galaxies of the Local Group $-$ are closer to the ones at the
early ages of the universe, when the peak of star formation occurred
\citep[$z\simeq 1.5$,][]{pei99}. Moreover, the lower dust-to-gas ratio in the Magellanic Clouds \citep{Koornneef82,Bouchet85}, in comparison with the Milky Way, assures a lower extinction, reducing the bias introduced by differential reddening, a well known limitation in the study of young stellar populations.

\begin{deluxetable*}{rcccccc}
\tablewidth{0pt}
\tablecaption{Sample from the photometric catalog of all stars found
in the region of LH~95 with HST/ACS imaging\label{t:catalog}}
\tablehead{
\colhead{}&
\colhead{R.A.}&
\colhead{DECL.}&
\colhead{$F555W$}&
\colhead{$\sigma_{\rm 555}$}&
\colhead{$F814W$}&
\colhead{$\sigma_{\rm 814}$}\\
\colhead{\#}&
\colhead{(J2000.0)}&
\colhead{(J2000.0)}&
\colhead{(mag)}&
\colhead{(mag)}&
\colhead{(mag)}&
\colhead{(mag)}
}
\startdata
1& 05~37~06.04& $-$66~21~37.15&  14.839&   0.001&  15.066&   0.001\\
2& 05~37~19.26& $-$66~21~07.88&  15.956&   0.001&  15.001&   0.001\\
3& 05~36~59.26& $-$66~21~20.66&  15.406&   0.001&  15.570&   0.001\\
4& 05~37~15.12& $-$66~21~44.39&  15.503&   0.001&  15.494&   0.001\\
5& 05~37~05.62& $-$66~21~35.39&  15.372&   0.001&  15.473&   0.001\\
6& 05~36~49.60& $-$66~23~26.16&  16.930&   0.003&  14.873&   0.001\\
7& 05~37~05.53& $-$66~21~59.65&  15.632&   0.001&  15.740&   0.001\\
8& 05~37~04.53& $-$66~22~01.02&  15.670&   0.001&  15.783&   0.002\\
9& 05~37~14.29& $-$66~22~52.00&  15.951&   0.001&  15.990&   0.001\\
10& 05~36~56.68& $-$66~21~10.40&  16.828&   0.001&  15.102&   0.001\\
11& 05~36~53.49& $-$66~21~55.55&  17.183&   0.002&  15.595&   0.001\\
12& 05~36~59.67& $-$66~22~02.10&  16.813&   0.001&  14.982&   0.001\\
13& 05~37~01.86& $-$66~22~35.65&  16.752&   0.002&  15.663&   0.001\\
14& 05~36~57.29& $-$66~21~49.03&  16.267&   0.001&  16.440&   0.001\\
15& 05~36~53.06& $-$66~22~11.28&  16.610&   0.001&  16.736&   0.002\\
\nodata & \nodata & \nodata & \nodata & \nodata & \nodata & \nodata
\enddata
\tablecomments{Magnitudes are given in the Vega system. Units of right
ascension are hours, minutes, and seconds, and units of declination are
degrees, arcminutes, and arcseconds. The spatial resolution of the ACS/WFC
is 0.05$\arcsec$ . }
\end{deluxetable*}

Most of the star formation occurs in dense cores of Giant Molecular Clouds, where one or more large stellar concentrations, known as OB
associations \citep{ambartsumian}, are formed. However, most of the star
formation studies in the Milky Way are carried out in smaller fields, like
the young Taurus, Lupus and Chamaeleon star-forming regions. Although it
has been shown that galactic OB associations host populations of faint,
low-mass PMS stars \citep{preibisch2002, sherry04, briceno2007} their
analysis is strongly limited by the contamination of background and
foreground evolved populations, requiring detailed measurements of proper
motions, time-consuming spectroscopy and multi-epoch photometry to define
their membership. On the other hand, the study of OB associations in the
LMC is less subject to contamination by field populations, due to the
limited distance spread \citep{caldwell86,cole98}. However, in order to
identify the faint red PMS stellar component of LMC associations,
photometry with instruments of high sensitivity and angular resolution is
required, due to the larger distance from us.

Recent findings from the {\sl Hubble Space Telescope} (HST) confirm the
presence of PMS stars in OB associations of the Magellanic Clouds.
\citet{gouliermis06a} studied this case in the LMC
association LH~52, using WFPC2 observations in the $V$- and
$I$-equivalent bands, and they discovered a population of objects in the color-magnitude diagram (CMD) consistent with T-Tauri stars. Subsequently, the {\sl Advanced Camera for Surveys} (ACS) has
been used to perform similar studies in other associations of the Magellanic Clouds,
such as NGC~346 \citep{nota06, gouliermis06b, hennekemper08,sabbi} and
NGC~602 \citep{schmalzl} in the SMC.  However, these studies could not
characterize the sub-solar mass function, because of the insufficient detection of such stars.

In the first part of our study ACS observations enabled us to discover a
rich  population of PMS stars in the star-forming region LH~95 in the
LMC \citep[from hereafter Paper I]{lh95first}. In this paper we focus on the IMF of these stars,
which is reliable for masses down to about $0.3$~M$_{\odot}$, considering
that our photometry allowed to perform precise completeness corrections
for stars with masses down to this limit. In \S~\ref{section:photometry}
we describe our photometry and the completeness of our data. In
\S~\ref{section:pmsstars} we apply the decontamination of the observed
stellar sample in the area of the association LH~95 from the contribution
of the general LMC field for the identification of the stellar members of
the system. We also investigate the spatial distribution of the identified
pre-main sequence population, in order to define the limits of the main
body of the system, and we apply extinction measurements. In
\S~\ref{s:pmstracks} we present our calculations for new PMS evolutionary
models in the observational plane. We make use of these models in combination
with our photometry for the construction of the system IMF of LH~95 in
\S~\ref{section:MF}. We discuss our findings in
\S~\ref{section:IMFdiscussion} and final conclusions are given in
\S~\ref{section:conclusions}.

\begin{figure}
\plotone{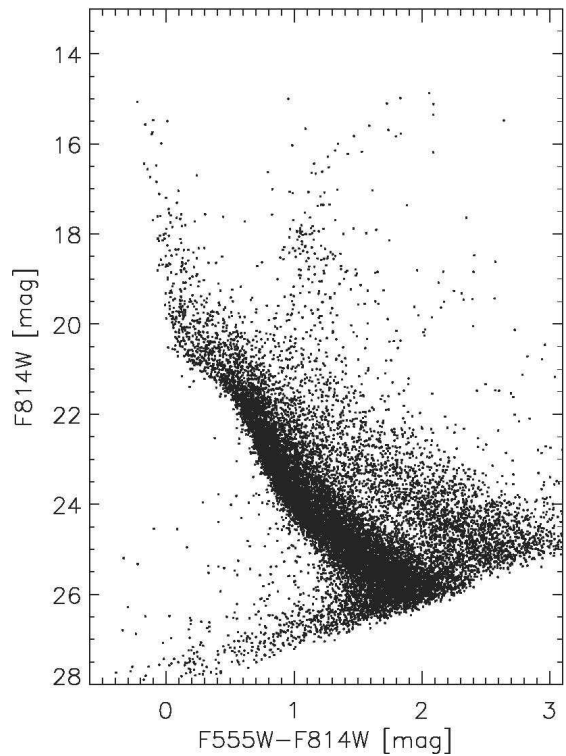}
\caption{$F555W-F814W$, $F814W$ Color-Magnitude Diagram of the entire
LH~95 region, according to the updated photometric catalog, which
includes about 900 stars more, recovered by visual inspection of the
sources rejected in the photometric study of Paper~I. It is worth noting that,
besides a dense main-sequence population, this CMD is characterized by a
remarkable number of PMS stars at faint magnitudes and red colors,
as indicated in Paper~I.\label{fig:CMD_total}}
\end{figure}


\section{Photometry}\label{section:photometry}

In this study we analyze the photometry presented in {\sl Paper I}, derived from deep observations with the
Wide-Field Channel (WFC) of ACS on-board {\sl HST} within the GO Program
10566 (PI: D. Gouliermis). Two pointings were observed, one centered on
the association LH~95 itself, which we refer to as the ``system'', and
another one about 2\arcmin\ to the west on an empty area,
typical of the local LMC field. We refer to the latter as the ``field''.
For each pointing observations in two photometric bands were available:
F555W and F814W, roughly equivalent to the Johnson V and Cousins I bands
respectively. The observations and their photometry are described in
detail in Paper~I.  We found about 16,000 stars in the system, about
2,600 of which lie in the PMS region of the CMD, and about 17,000 in the
field.

\begin{figure*}[ht!]
\epsscale{1.1}
\plottwo{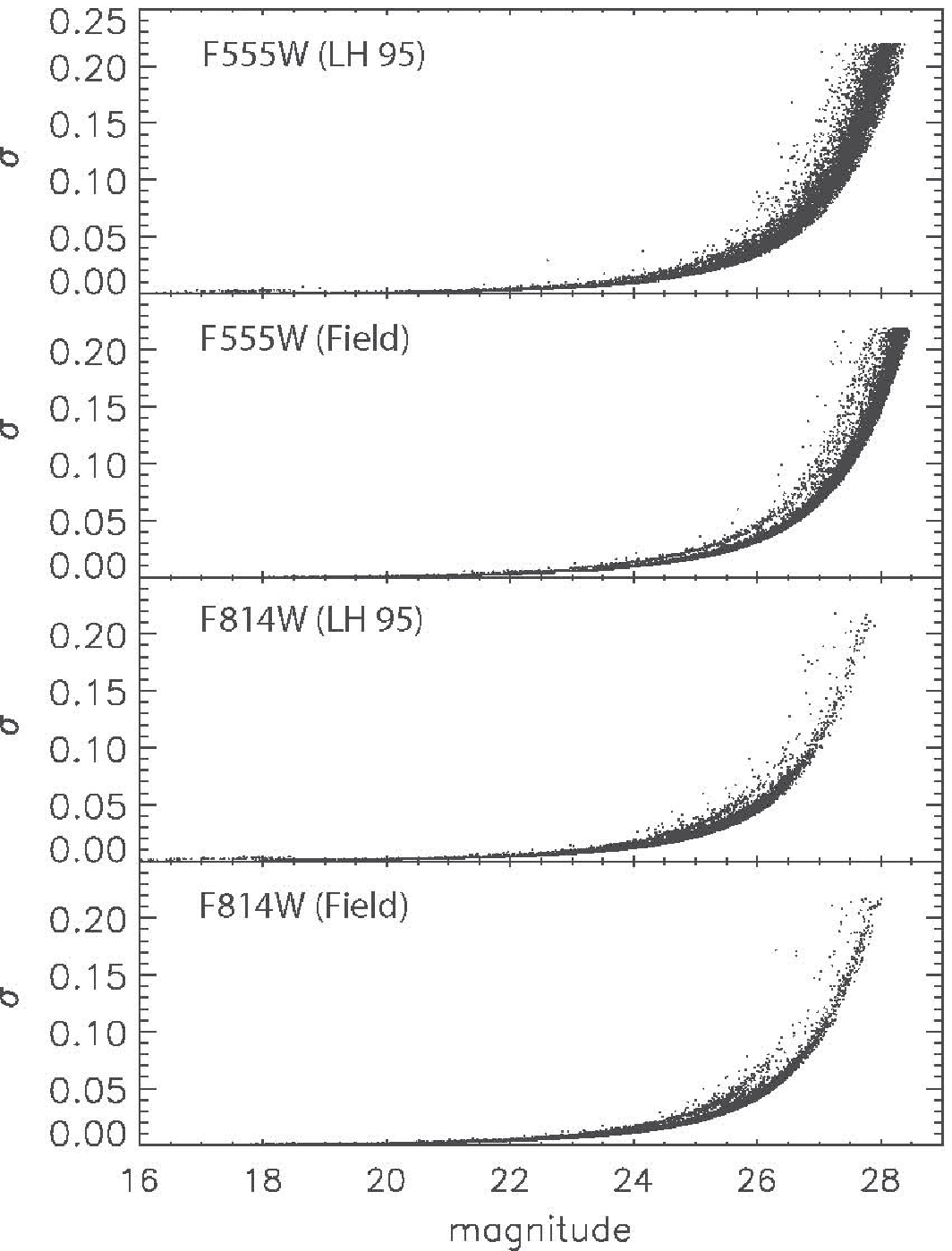}{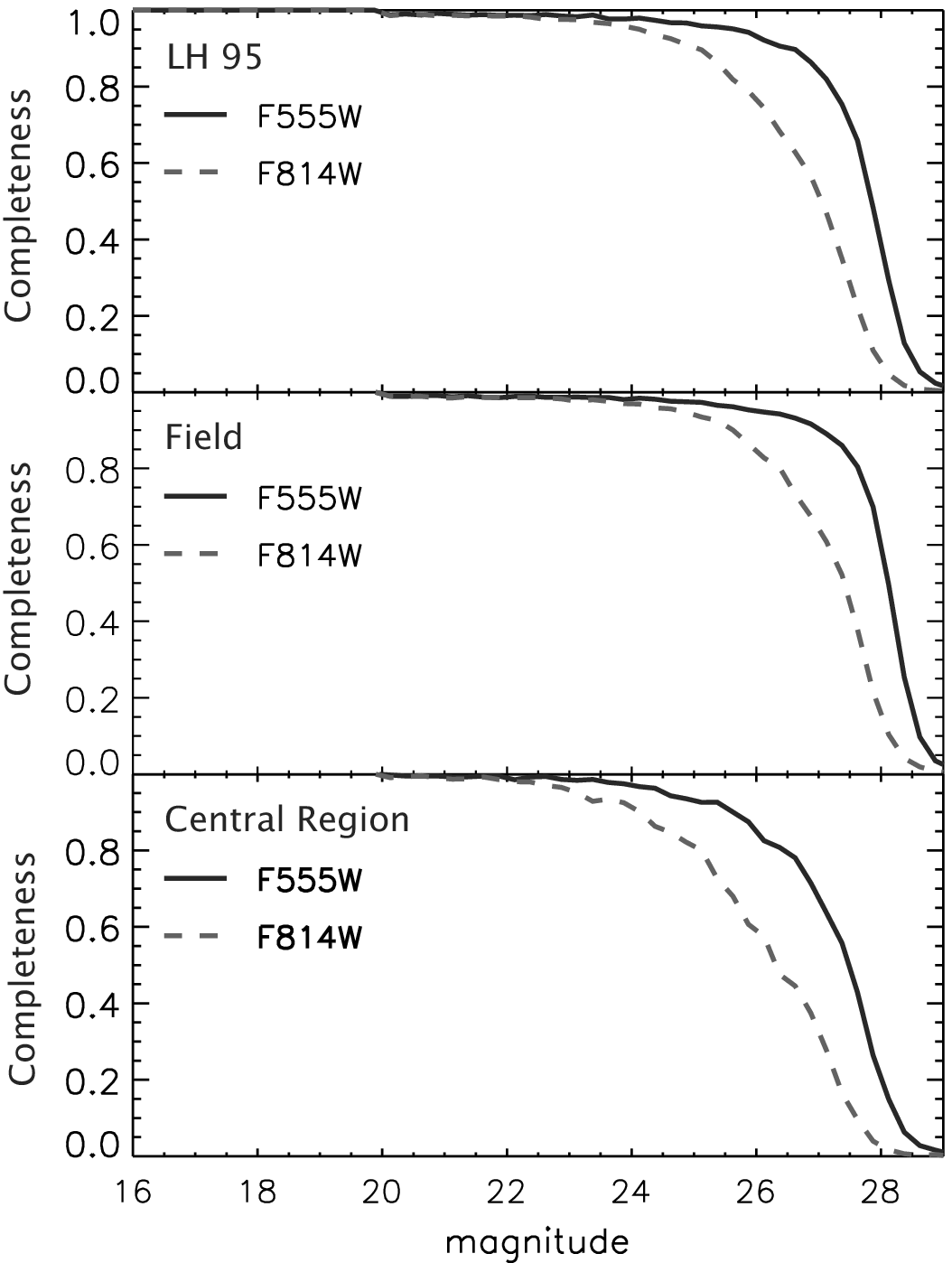}
\caption{{\sl Left panel}: Distribution of the photometric errors, for
both the F555W and F814W band for the LH95 system and the LMC
field. {\sl Right panel}: Completeness of our photometry in the entire
observed areas of both the system (top) and the field (middle), as well
as for the {\sl central region} of the association LH~95 (see
\S~\ref{section:centralregion}).\label{fig:photomerrors_completeness}}
\end{figure*}

In the present work we enhance the photometric catalog of stars derived
in Paper~I by searching for eventually missing objects, especially in the low-mass
PMS regime. Photometry was obtained using the ACS module of the package
DOLPHOT\footnote{DOLPHOT is an adaptation of the photometry package {\tt
HSTphot} \citep{Dolphin2000}. The software and its documentation can be
obtained from {\tt http://purcell.as.arizona.edu/dolphot/}.} (Ver. 1.0). The selection of stars in our original photometry was based on the
quality parameters estimated for each detected source by the package.
However, objects which are actual stars may have been rejected as
spurious detections in crowded regions, especially in the presence of a
non-uniform background due to nebular emission, as well as in the
neighborhood of very bright sources. Therefore, we considered the
catalog with all the rejected sources and we performed a visual
inspection of each of them on the original ACS FITS images constructed
with {\tt Multidrizzle}. With this process we recovered about 900
additional stars. The updated photometric catalog of the association
LH~95 includes in total 17,245 stars. A sample of this catalog is shown
in Table~\ref{t:catalog}. The corresponding CMD is shown in Fig.
\ref{fig:CMD_total}. Typical uncertainties of our photometry as a
function of the magnitude for both filters are given in
Figure~\ref{fig:photomerrors_completeness} (left), for both the
``system'' and the ``field'' ACS pointings.

\subsection{Completeness}
\label{section:completeness}

The completeness of the data is evaluated by artificial star experiments
with the use of lists of almost 400,000 artificial stars created with
the utility {\sl acsfakelist} of DOLPHOT for each of the observed areas
of the system and the field. This utility enables to add stars of given
magnitudes to all the frames simultaneously, searching for a detection
in none, one, or both photometric bands. In this way it is possible to
derive automatically a 2-dimensional completeness function for both
color ($F555W-F814W$) and magnitude $F555W$ or $F814W$. Considering that
at any given faint magnitude in one band the color term covers a range
of up to $2$ magnitudes, as shown in the CMD of
Figure~\ref{fig:CMD_total}, allowing the magnitude in the other band and
therefore the completeness to vary significantly, this two-dimensional
approach is necessary for an accurate treatment of the completeness of
our data. Therefore, we define with our artificial star technique the 2D
completeness as a function of ($F555W-F814W$) and $F814W$, by computing
the ratio of the detected sources over the total added stars in a grid of
equally spaced bins $0.25$ magnitudes wide in ($F555W-F814W$)
and and $0.5$ magnitudes in $F814W$.

Such completeness maps have been constructed for both the system and the field,
enabling, via interpolation, an accurate determination of the actual
completeness at any given point of the CMD. The right panel of
Figure~\ref{fig:photomerrors_completeness} shows the completeness function
in both bands for both the system and the field (top and middle
respectively). The completeness is found to vary with location, and
therefore different regions in the observed field-of-view are characterized by
different completeness functions according to their crowding. In the
bottom panel of the figure we show the completeness as a function of
magnitude for the central region of the association, as it is defined in
\S~\ref{section:centralregion}. It is remarkably lower than that in
the entire field-of-view, due to higher crowding of faint stars and
over-density of the bright ones.

To avoid further contamination by spurious objects in our
catalog, which could have been included through additional stars, we trimmed
the data set used in this work excluding sources with $V>28.5$~mag or
$I>28$~mag. At these magnitude limits stars are faint enough to be well
below any reasonable completeness limit. The completeness in the
``system'' CMD at V=$28.5$ mag and I=$28$ mag is less than 1\%. Consequently, their removal does not affect the
quality and completeness of our data.


\section{The Pre-Main Sequence Population of
LH~95}\label{section:pmsstars}

 \subsection{Field subtraction}
 \label{section:fieldsubtraction}

The observed stellar population in the area of the association LH~95 is
naturally contaminated by the general field population of LMC.
Therefore, in order to study the stellar members of the association
alone, and in particular its PMS stars, the field contamination should
be removed. In our case, having at our disposal photometry in only two
bands, the possibility to distinguish if a given star is a true member
or not is limited, but a statistical approach can provide accurate
results.

We use a Monte Carlo technique, which
considers that a star belongs to the cluster or the field with a
probability, which depends on the local density in the same part of each
CMD (of system and field). In order to maximize the statistical accuracy
of the result, two iterations are applied. First, the field subtraction
is applied to the entire observed field-of-view centered on the system,
enabling to study the spatial distribution of the LH~95 stars
and to isolate the boundaries of the association. After that, the
study is limited on this subregion of the frame, and another field
subtraction is carried out on the central region providing an accurate
estimate of the membership of the sources included.

Specifically, for the first iteration we consider each star in the
catalog of the system and we select an elliptical region in the CMD of
the system centered on the position of this star, and the same region in
the CMD of the field. Each such region is being chosen with semi-axes
$\Delta I = 0.5$~mag and $\Delta (V-I) = 0.17$~mag. This particular size
has been chosen as to be wide enough to include a considerable number of
stars, and still small enough to achieve a considerably high resolution
in the resulting CMD. We compute the probability of each star in the area
of the system to belong to the field as: \begin{equation}
P=A\cdot\frac{N_{\rm fld}}{N_{\rm sys}}\cdot\frac{C_{\rm sys}}{C_{\rm
fld}} \label{equation:mc} \end{equation} where $N_{\rm sys}$ and $N_{\rm
fld}$ are the numbers of stars included in the elliptical region of the
CMD of the system and the field respectively. $C_{\rm sys}$ and $C_{\rm
fld}$ are the completeness factors, for system and field, measured for
exact position of the considered star in the CMD, as explained in
\S~\ref{section:completeness}. $A$ is a normalization factor which takes
into account differences in the field stellar density between the two
regions. From repeating measurements in different parts of the area of
the field and that of the system we found that on average the field includes
systematically $\sim 1.3$ times more main sequence stars than LH~95. Therefore, considering that both areas cover equal surfaces, we
set $A=0.77$. According to Eq.~\ref{equation:mc}, when $P\geq1$ the star
is considered to be a field star and it is removed; $P=0$ means that
there are no field stars in the field CMD in the elliptical neighborhood
of the star and therefore it is considered a member of the association.
When $0<P<1$ the star is randomly kept or rejected as a member star with
a probability equal to $P$.

\begin{figure}
\epsscale{1.1}
\plotone{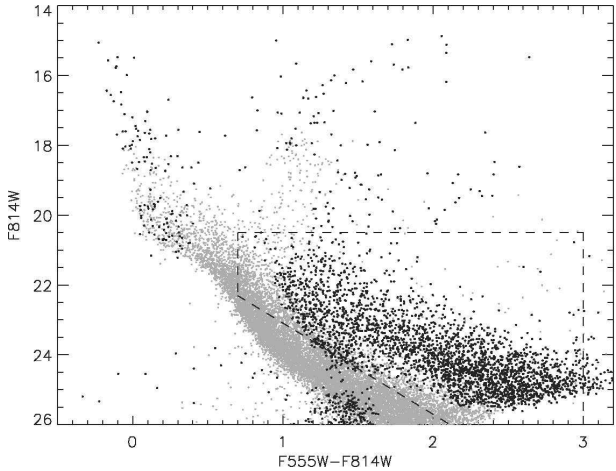}
\caption{$F555W-F814W$, $F814W$ Color-Magnitude Diagram (CMD) of the
entire observed area around LH~95. The stars which are identified as
members of the system alone and not of the field of the LMC based on the first field-subtraction iteration (see
\S~\ref{section:fieldsubtraction}) are plotted with thick symbols. The
dashed line sets the limits of the loci in the CMD of the PMS alone. The
spatial distribution of these stars has revealed the existence of young
stellar sub-groups within the association (Paper~I; see also
\S~\ref{section:centralregion}).
\label{fig:CMD_first_bkgsub}}
\end{figure}

The result of the first field-subtraction iteration, applied to the
entire LH~95 field of view is shown in
Fig.~\ref{fig:CMD_first_bkgsub}, where stars flagged to belong to the
association are plotted by black dots. As shown in Paper~1, it is
evident that the PMS population belongs entirely to LH~95, while all
lower main sequence stars turn out to belong to the LMC field. The upper
main sequence (UMS) shows an overabundance of stars in the area of the
system, demonstrating the presence of the corresponding population in
the association itself.

\subsection{Topography of the PMS Stars}
\label{section:centralregion}

In this section we study the spatial distribution of the PMS stars in
the area of LH~95, as an indication of the places where star formation
occurs. For this purpose, we construct surface density maps of the
whole area by counting the stars identified as system members after
applying the Monte Carlo removal of the field contamination as described
in \S~\ref{section:fieldsubtraction}. Since we are interested in the PMS
cluster properties, we considered only the PMS population, as selected
in the region of the CMD shown in Figure~\ref{fig:CMD_first_bkgsub}.

The star counts are performed by dividing the whole ACS field-of-view
into a grid of $50 \times 50$ elements, each corresponding to a size of
$\sim 85$ pixels ($\simeq 1$pc), and by counting the (field-subtracted)
PMS stars in each of them. The selection of the specific grid-element
size, which corresponds to the ``resolving efficiency'' of the detection
of stellar concentrations, was chosen, after several experiments, as the
most appropriate for revealing the smallest concentrations with a
physical meaning.

\begin{figure*}
\epsscale{1.15}
\plottwo{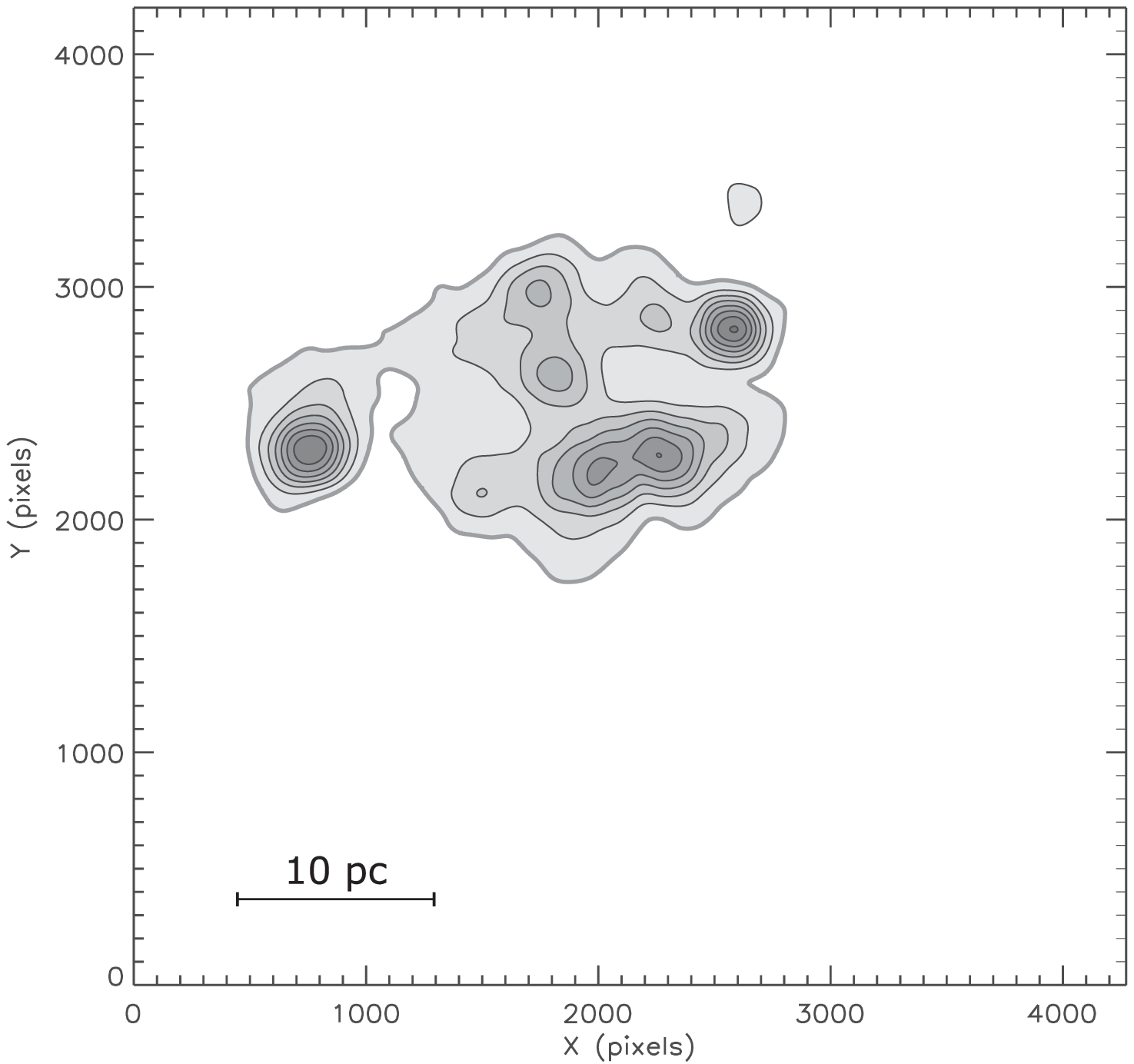}{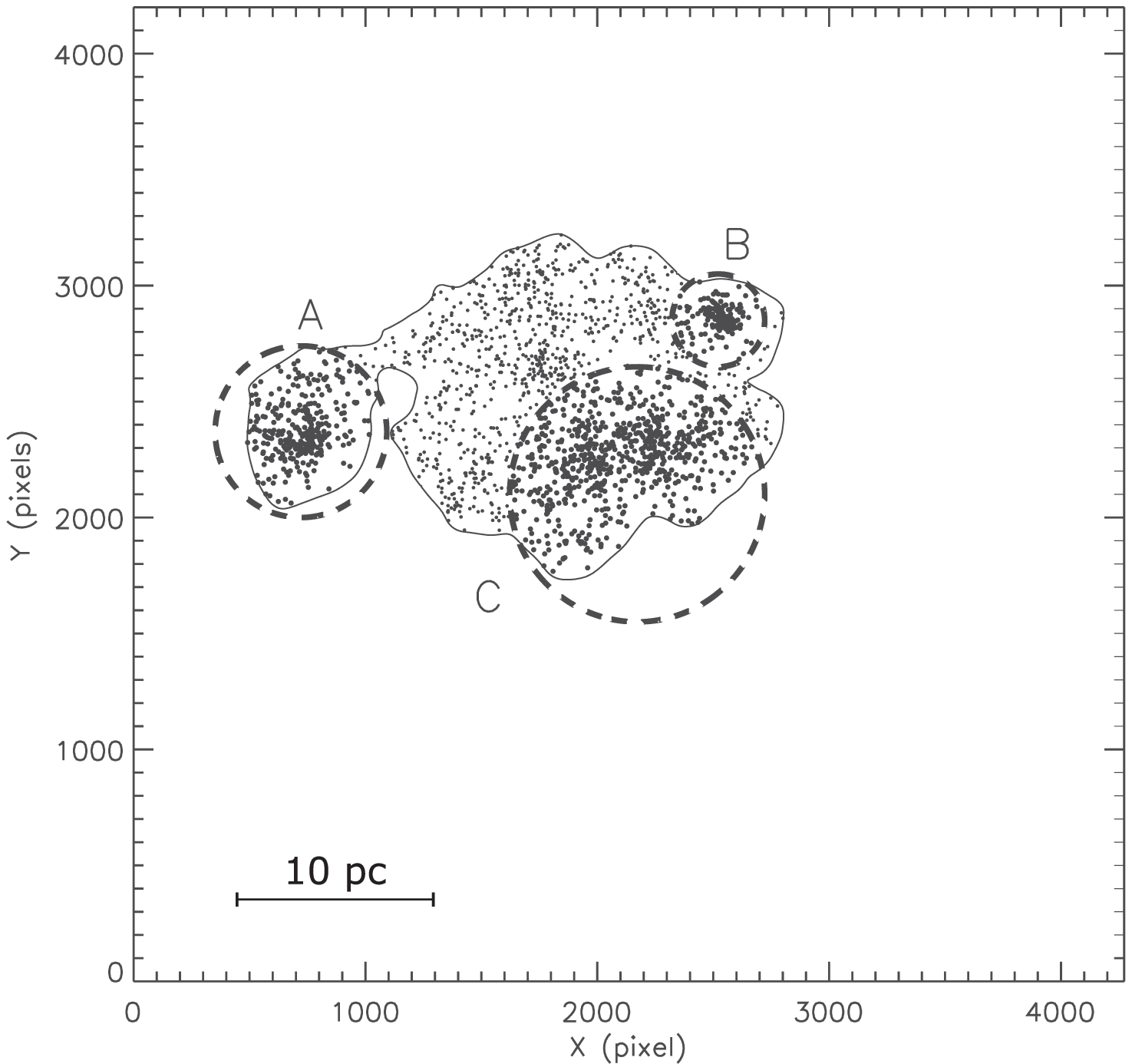}
\caption{{\sl Left panel:} Isodensity contour map of the area of the
system constructed from star counts of the PMS stars alone. Isopleths
are drawn in steps of 1$\sigma$, where $\sigma$ corresponds to the
standard deviation of the background density, starting from the
3$\sigma$ level. This level defines the limits chosen for the selection
of the stars used for the construction of the CMD of the central part of
the area of the system shown in Figure~\ref{fig:CMD-sub} (left and right
panels). This map demonstrates that, as we found in Paper~1,
PMS stars are concentrated in few compact subgroups within the
association. These density peaks in the PMS stars coincide with those of
the Upper Main Sequence (UMS) stars (Paper~I), suggesting that these subgroups contain fully
populated IMFs. Units are in pixels on the ACS drizzled frame. {\sl
Right panel:} Positions of the three most prominent concentrations
(subclusters) of PMS stars, which appear as the densest substructures of
the main part of the association LH~95. \label{fig:densitymap}}
\end{figure*}

In order to remove from this map eventual noise, we applied a smoothing
on the density map with a kernel of size $\sim 2.5$pc $\simeq 200$
pixels. The majority of the PMS stars is found to be located within a
compact region $\sim 1\arcmin \times 2\arcmin$, with a density well
above the threshold of $3\sigma$, where $\sigma$ is the standard
deviation of the background surface density. The derived two-dimensional
density map is shown in Figure~\ref{fig:densitymap} (left), where the
$3\sigma$ density limit, corresponding to the first (lower) isopleth, is
chosen to define the statistically significant concentration of the PMS
population of the association LH~95.

\begin{deluxetable*}{crrrrccccr}
\tablewidth{0pt}
\tablecaption{Characteristics of the {\sl Central Region} (LH~95) and its
three subclusters \label{t:subcluster}}
\tablehead{
\colhead{}&
\colhead{RA}&
\colhead{DEC}&
\colhead{Size}&
\colhead{$r_{\rm h}$}&
\colhead{$M_{\rm tot}$}&
\colhead{$\varrho$}&
\colhead{$t_{\rm d}$}&
\colhead{$t_{\rm cr}$}&
\colhead{$t_{\rm relax}$}\\
\colhead{}&
\colhead{(J2000.0)}&
\colhead{(J2000.0)}&
\colhead{(pc)}&
\colhead{(pc)}&
\colhead{(10$^3$~M{\solar})}&
\colhead{(M{\solar}~pc$^{-3}$)}&
\colhead{(100~Myr)}&
\colhead{(Myr)}&
\colhead{(Myr)}
}
\startdata
LH~95 & 5~37~06.50 & $-$66~22~03.50 & 33.5 & 6.5 & 2.37 & 0.47 & 0.9 & 6.4 & 433.0 \\
A     & 5~37~15.25 & $-$66~21~40.87 &  8.7 & 1.7 & 0.33 & 1.28 & 2.4 & 2.2 &  26.9 \\
B     & 5~36~59.78 & $-$66~21~36.88 &  4.7 & 0.8 & 0.19 & 2.70 & 5.1 & 0.9 &   6.7 \\
C     & 5~37~03.96 & $-$66~22~09.73 & 12.9 & 3.0 & 0.98 & 1.11 & 2.1 & 3.1 &  95.5 \\
\enddata
\end{deluxetable*}

From here on, we refer to the region confined by the $3\sigma$ density
isopleth of Figure~\ref{fig:densitymap} as the {\sl central region} of
LH~95, and we focus our subsequent analysis on the stellar population
included within the boundaries of this region, as the most
representative of LH~95. In Figure~\ref{fig:densitymap}~(left) can be
seen that within the central region there are smaller substructures
characterized by a higher projected density of PMS stars. We select the
three most prominent ones, isolating circular areas within the central
region as shown in Figure~\ref{fig:densitymap}~(right). We refer to
these substructures as {\sl subcluster} {\sl A}, {\sl B} and {\sl C}
respectively. The positions and sizes of the selected circular regions
around the subclusters, shown in Figure~\ref{fig:densitymap}~(right),
are given in Table \ref{t:subcluster} with those for the whole central
region. We measure the total mass, $M_{\rm tot}$, included in each
stellar concentration, assuming that all systems follow a mass function
similar to the standard Galactic field IMF \citep{kroupascience}, and
extrapolating their stellar content down to 0.08~M{\solar}. We then
estimate the stellar density, $\varrho$, and the disruption time,
$t_{\rm d}$, of each cluster following the method by \citet[][their
section 6]{gouliermis02}.
The latter is given as \citep{spitzer58}: \begin{equation} t_{\rm d} =
1.9 \times 10^{8} \varrho \bigg( \frac{{\rm M}_{\odot}}{{\rm
pc}^{3}}\bigg)~~{\rm years}~.  \end{equation} Moreover, the dynamical
status of a stellar system is defined by two additional time-scales, the
{\sl crossing} and the {\sl two-body relaxation} time \citep{kroupa08},
which are given as: \begin{equation} t_{\rm cr} \equiv \frac{2 r_{\rm
h}}{\sigma}~~~{\rm and}~~~t_{\rm relax} = 0.1 \frac{N}{\ln{N}} t_{\rm
cr} \label{eq:tcr}\end{equation} respectively. The three-dimensional
velocity dispersion of the stars in the cluster, $\sigma$, is given as
\begin{equation} \sigma = \displaystyle{ \sqrt{ \frac{{\rm G}M_{\rm
tot}}{\epsilon r_{\rm h}}}~~,} \end{equation} where $\epsilon$ is the
star formation efficiency (SFE) and $r_{\rm h}$ the half-mass radius of
the cluster. In order to make a rough estimation of the aforementioned
time-scales for the subclusters and the whole central region of LH~95,
we apply the formulas of Eq.~\ref{eq:tcr}.

The SFE in several nearby Galactic gas-embedded clusters has been found to
range typically from 10\% to 30\% \citep{lada03} or 20\% to 40\%
\citep{kroupa08}, value which increases with time while the gas is
removed. However, the average age of the system is 4~Myr (see
\S\ref{section:MF}), slightly greater than the typical time necessary to
remove most of the gas \citep{lada03}. Therefore, and taking into account
the very low optical extinction (see \S\ref{section:centralregion}), LH~95
should be considered as an emerging cluster, meaning that it should be at
the process of separation from the parental cloud. Furthermore, according
to \citet{wilking83}, a high SFE is required for a bound young cluster to
emerge from its parental cloud. As a consequence, we consider a value of
$\epsilon \simeq 0.4$ as more adequate for our system. We derive, thus,
values of $\sigma$ of the order of 2 km~s$^{-1}$ for our objects and we
provide the additional estimated structural parameters for each system
also in Table~\ref{t:subcluster}. We find that the crossing time for the
whole region is greater than the age of the system, and this is quite
consistent with the observed sub-clustering of the system.

From a comparison between the characteristics of the subclusters and the
central region, as they are given in Table~\ref{t:subcluster}, one can
see that there is a significant fraction of mass outside the immediate
regions of the subclusters. This can also be seen in the maps of
Figure~\ref{fig:densitymap}, where a prominent population of PMS stars
is easily distinguished away from the subclusters, spread in the area
among them within the central region. We estimate that the fraction of
{\sl distributed} stellar mass over the total corresponds to
$\sim$~40\%, with the remaining 60\% being {\sl clustered}.  This
fraction of distributed PMS stars in LH~95 is consistent with
statistical analyses of several Galactic nearby star-forming regions
that predict \lsim~60\% \citep[see review by][]{allen07}. However, it is
not clear if this distributed population is the result of the star
formation or a merging process. Detailed simulations do predict that
subclusters may merge to form a larger one \citep[e.g.][]{fellhauer06},
but our data do not allow us to verify if this takes place in LH~95. In
any case, the stellar density and disruption time of all considered
systems, including the whole central region, show that all are rather
compact and none of them is under disruption. Subclusters A and B,
located at the east and west of the central region, are particularly
compact and roughly spherical shaped, while subcluster C, close to the
southern limit of the central region, presents a more elongated and
extended distribution of PMS stars. This subcluster corresponds to the
visibly prominent main part of the association (see Figure~1 in
Paper~I).

\begin{figure*}
\plotone{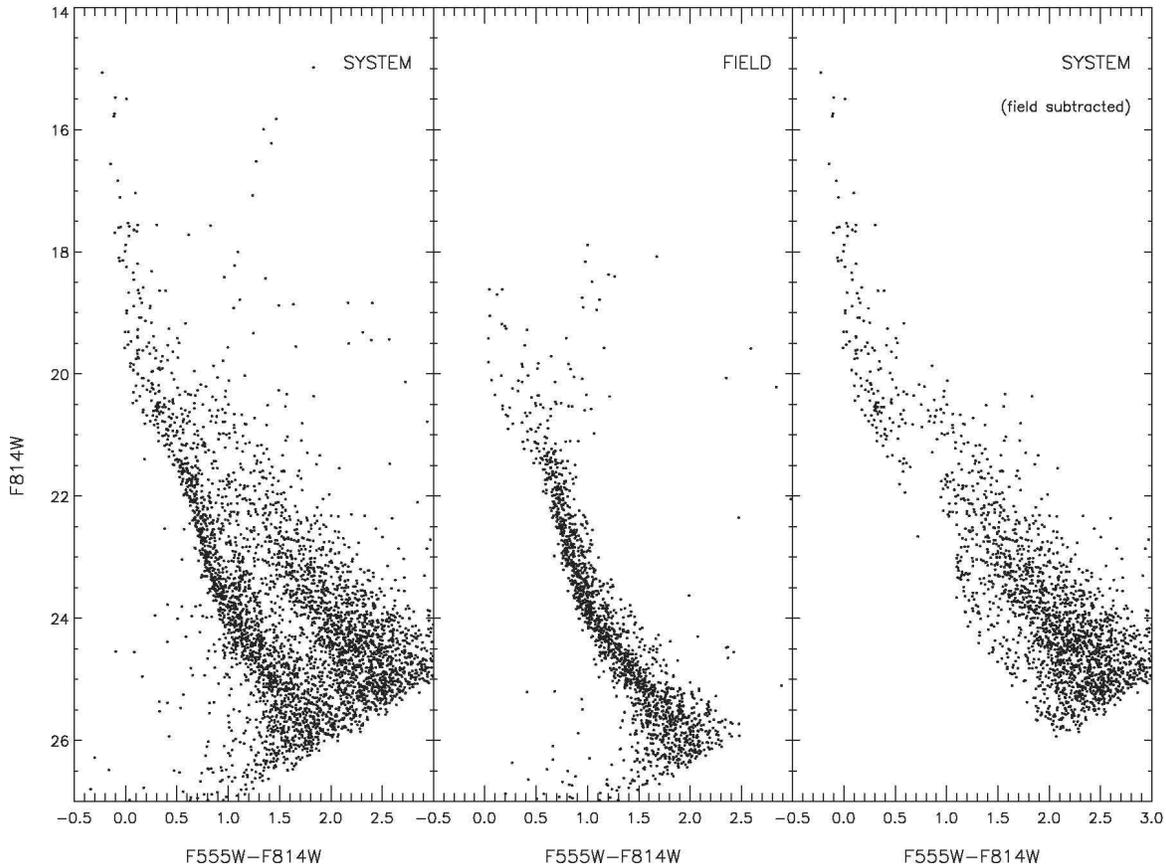}
\caption{Left: CMD of all stars observed in the central part of the area
of the system. This part was selected as the most representative of the
association based on star counts (\S~3.2; Figure~\ref{fig:densitymap}). Middle: CMD
of the corresponding part of the area of the field, which was used for
the decontamination of the CMD of the system from the contribution of
the field stellar population. Right: CMD of central area of the system
after the field subtraction was applied with the use of an advanced
Monte Carlo method (\S~3.1). This CMD highlights the UMS and PMS stars
as the prominent stellar populations of the association LH~95,
demonstrating its youthfulness.\label{fig:CMD-sub}}
\end{figure*}

\subsection{LH~95 Central region}
\label{section:fieldsubtractio2}

For the detailed study of the PMS population in LH~95, while minimizing
the contamination by other stars, we focus on the central region of the
system.
Specifically, we consider the entire population included in the central region and we apply, for a second time, the Monte Carlo technique for field subtraction, taking into account the ratio between the surface of the observed LMC field and that of the {\sl central region}.

It should be noted that
this second application of the field subtraction technique to the central
region is not equivalent of simply isolating the stars, which are found to
be located within the boundaries of the region after the first iteration
of the Monte Carlo subtraction.

Naturally, the central region has a higher density of system stars
(specifically PMS stars), and therefore a lower relative contamination by
field stars. Consequently, the actual fraction of stars which are marked
to be field stars is lower, reducing the probability of (statistically
probable) wrong membership assignments. The
number of field stars included in the elliptical region of the CMD $N_{\rm
fld}$ is computed based on the photometry in the entire field and scaled
according to the area occupied by the LH~95 central region, which we found
to be $\sim~12.6\%$ of the total ACS field-of-view.

The result is shown in Figure \ref{fig:CMD-sub}, where we plot the CMDs of
the whole population included within the boundaries of the central region
(left panel), the one of the whole area of the field, sampled randomly and
down-scaled according to the area coverage of the central region (central
panel), and the corresponding CMD of the central region after the field
contribution has been removed (right panel). From this figure it is evident
that the observed stellar population of the low main sequence (LMS) can be
entirely considered as field population. On the other hand the association
has a prominent upper main sequence (UMS) component, and all pre-main
sequence stars, easily distinguished by redder colors, are solely present
in LH~95 alone. Furthermore, the second application of our field
subtraction technique, limited to the central region, allows us to reduce
considerably the contamination by objects not located in the PMS and UMS
parts of the CMD, which are still present as residuals from the first
application of the method on the entire field-of-view
(Figure~\ref{fig:CMD_first_bkgsub}).

In the CMDs of Figure \ref{fig:CMD-sub} can be noted that the field
population has a tighter sequence than the PMS population, which may
indicate an age-spread among the PMS stars of LH~95 (see also
Figure~\ref{fig:CMDwithtracks}). Indeed, the locations of low-mass PMS
stars in the CMD of young stellar systems of our Galaxy often show a
widening, which may be evidence for an age-spread in the system
\citep[e.g.][]{palla00}. However, several characteristics of these PMS
stars (being T~Tauri stars), such as variability and circumstellar
extinction, can cause considerable deviations of
their positions in the CMD \citep{sherry04}, which may be misinterpreted
as an age-spread. As a consequence, detailed simulations of the
characteristics of PMS stars are required to quantify the effect of
these characteristics on their positions in the CMD and to conclude on
any true age-spread \citep{hennekemper08} with the use of
population-synthesis techniques.

In the following sections we focus our study on the isolated, field
subtracted population in the LH~95 central region, shown in
Figure~\ref{fig:CMD-sub} (right).

\subsection{Interstellar Extinction}\label{section:reddening}

With photometry available only in two bands it is not possible in
general to measure the visual extinction $A_{V}$ independently for every
star of the system. It is, however, possible to perform a statistical
approximation obtaining the average value of reddening as well as its
distribution. For this purpose we consider the UMS stars ($I\leq 20.0$ mag)
in the central part of LH~95, after performing field subtraction, and we
compare the position observed in the CMD to that expected according to
the Padova grid of evolutionary models in the ACS photometric system
\citep{girardi02}. Considering that LH~95 is a very young association
($\tau$~\lsim~10~Myr) we assume that all UMS stars have a age equal to
the youngest available isochrone in the Padova models, $\log{\tau}=6.6$
for a metallicity of $Z=0.008$ (typical value for the LMC;
\citealt{kontizas}) and distance modulus $m-M=18.41$~mag, compatible
with several distance measurements for the LMC, available in the
literature (e.g. \citet{alcock2004}).

\begin{figure*}
\epsscale{1.1}
\plotone{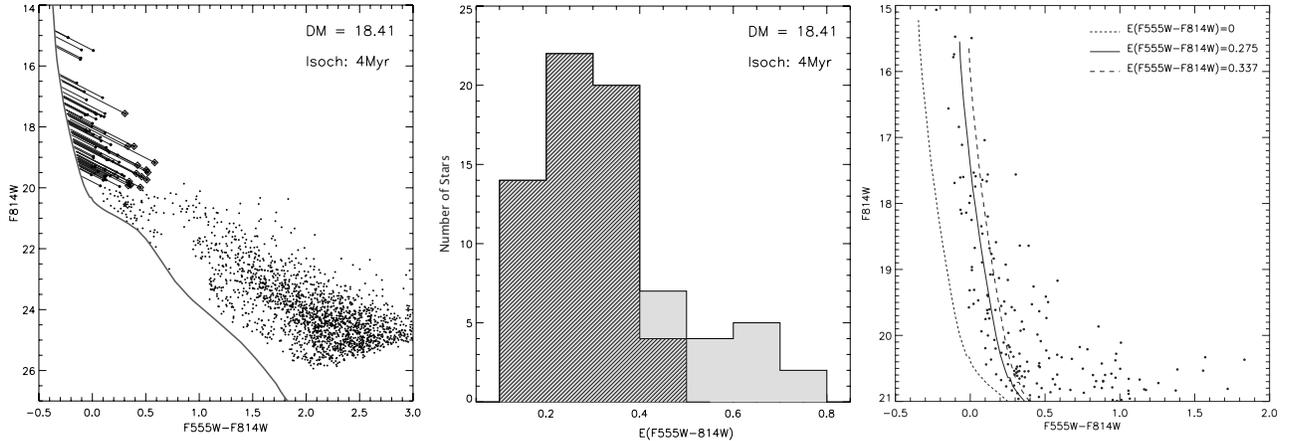}
\caption{Determination of the reddening in the region of LH~95.
{\sl Left:} The field subtracted stellar population within the central
area of LH~95 with the isochrone model for $\log_{10}{\tau}=6.6$ and the
corresponding individual reddening vectors of the upper main sequence
stars overlayed. Square symbols represent objects with
$(V-I)>0.3$~mag, which may be considered as candidate Herbig Ae/Be
stars, due to their highly reddened colors. {\sl Middle}: Distributions
of the measured reddening for all UMS stars (filled histogram) and
excluding candidate Herbig Ae/Be (darker histogram). {\sl Right}: UMS
CMD showing the Girardi et al. (2002) isochrone for $E(F555W-F814W)=0$
and with the mean reddening from the distributions shown in the central
panel applied. \label{fig:red}}
\end{figure*}

For every star we compute the intersection between the reddening vector
and the isochrone, and we obtain, thus, the reddening value in terms of
$E(F555W-F814W)$. Since the ACS photometric system is slightly different
from the standard Johnson-Cousins system, we compute the exact reddening
parameters for our ACS bands using the typical galactic extinction law
of \cite{cardelli}, parameterized by a value of $R_{V}=A_{V}/E(B-V)=3.1$.
We consider a set of template spectra taken from the \textsc{NextGen}
catalog \citep{nextgen}, and we apply the extinction curve with an
arbitrary $A_V$ which we choose to be equal to 1, and we measure the
extinction $A_{F555W}$ and $A_{F814W}$ by means of integration of the
original and reddened spectra within the ACS filter profiles. In this way we compute $R_{F555W}=A_{F555W}/E(F555W-F814W)\simeq2.18$ and
$R_{F814W}=A_{F814W}/E(F555W-F814W)\simeq1.18$.

The derived reddening distribution, shown in Figure~\ref{fig:red} ({\sl
middle}), appears to have a tail extended to higher values (filled
histogram). This tail appears due to a number of UMS objects, which
show a high color excesses. Taking into account all objects, including
the ones with high color excess, the average reddening has a value
$E(F555W-F814W) \simeq 0.34$~mag, corresponding to an optical extinction
of $A_{V}=1.56$. However, these red objects are being suggested to be
Herbig Ae/Be (HAeBe) stars \citep{gouliermis02}. Such stars of
intermediate mass, aged between 1 and 10~Myr, occupy the faint end of
the UMS with $17.5<F814W<20$~mag. Being in their pre-main sequence
phase, they have an intrinsic color higher than that of MS stars in the
same luminosity range. Consequently, if indeed these are not MS but
HAeBe stars, which is possible due to the youthfulness of the system,
then the derived reddening would be biased towards higher values.

We estimate the mean reddening without considering these stars (darker
histogram in Figure~\ref{fig:red} - {\sl middle}) and we derived a lower
mean value of $E(F555W-F814W) \simeq 0.275$~mag, corresponding to an
optical extinction of $A_{V}=0.6$ mag. In Figure~\ref{fig:red} ({\sl right})
both values are applied to the 4~Myr isochrone model (the youngest
available in the grid of models by the Padova Group). From this figure it is evident that the difference between the two values is not dramatic,
producing more or less the same shift of the model in the CMD. As a
consequence, and in order to avoid any biases due to the presence of
young stellar objects in LH~95, we consider an optical reddening of
$E(F555W-F814W) \simeq 0.3$~mag as the most representative for the
region. We repeated the construction of the reddening distributions
shown in Figure~\ref{fig:red} (middle) and the determination of the mean
reddening values using different isochrones, spanning an age range
between 4 and 7~Myr and assuming different distances varying by 0.2
mag around the assumed distance modulus. We found that the mean
reddening remains practically unchanged within 0.02 magnitudes, a
difference small enough to confirm the accuracy of our estimation.

\section{New set of observational PMS models}\label{s:pmstracks}

In the last decades several theoretical models for the pre-main sequence
stellar evolution have been computed. Some of the most popular, for PMS
masses above the hydrogen burning limit, are the models of
\cite{siess2000}, \cite{pallastahler}, \cite{dantona94,dantona97} and
\cite{swenson94}. These models are generally expressed in terms of
physical quantities, such as the effective temperature and the total
bolometric luminosity, describing the evolution of stars in the HR
diagram. However, one of the primary aims of stellar evolution theory is
the explanation of the observed photometric data of stars in order to
extract their masses and ages from their magnitudes and colors, and
therefore a conversion between physical and observable quantities of the
models is required. Such a conversion is specifically described by
\cite{siess2000}, who include the transformation between $T_{\rm eff}$
and $L$ to colors and magnitudes in the $UBVRI$ Cousins system and
$JHKL$ infrared bands for their models. These authors use simple
relations $T_{\rm eff}$ versus color, as well as bolometric corrections
from either \cite{siess97} or \cite{kenyon}. These conversion tables,
derived from observations of stellar clusters, are valid for solar
metallicity dwarf stars, but the discrepancies can be significant when
one deals with populations with different stellar parameters. In
particular, the age of a star is related to the stellar radius and
therefore to the surface gravity, which introduces differences in the
spectral behavior, and therefore a population of different ages or
metallicities may require different conversion relations. This issue can
be important for e.g. cold M-type stars, for which the broad molecular
absorption bands dominate the optical spectra, affecting integrated
colors and color corrections.

\begin{figure}
\plotone{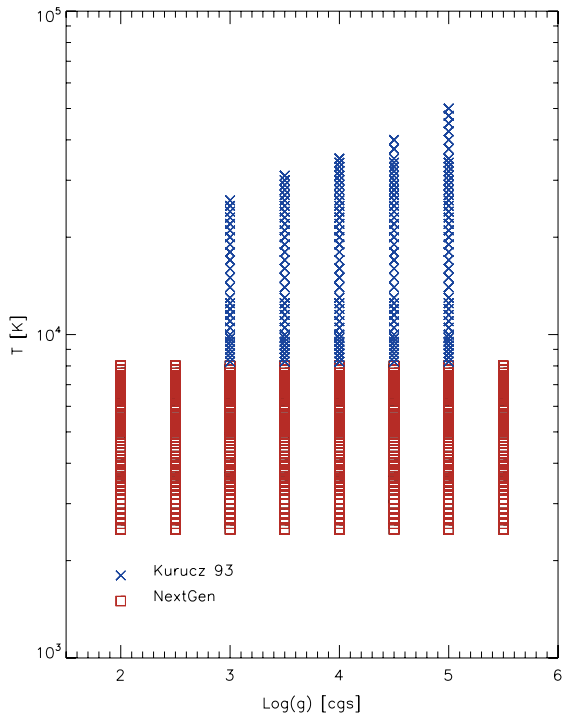}
\caption{Distribution in the $T_{\rm eff}$, $\log{(g)}$ space of the
synthetic spectra collected for our conversion of theoretical
evolutionary models into the observable plane. The same spectra are used
for metallicities of $[M/H]=0,\ -0.5,\ -1,\ -1.5$. Squares are spectra
taken from the \textsc{NextGen} models of \cite{nextgen}, while
crosses are those of \cite{kurucz79, kurucz93} (see
\S~\ref{s:pmstracksmethod}).
\label{fig:spectra}}
\end{figure}

A more thorough method to analyze this issue is to make use of synthetic
atmosphere models, performing photometry directly on synthetic spectra.
\cite{girardi02} applied this method for their evolutionary models for
evolved populations, which were converted in several photometric systems
using a grid of atmosphere models described by three parameters
($[M/H]$, $\log{(T_{\rm eff})}$, $\log{(g)}$). This method was also
applied for stars of sub-solar mass and brown dwarfs by
\cite{bcah98,bcah}. For low- and intermediate-mass stars, however, there
are no similar conversions of evolutionary models into the observable
plane for various photometric systems available. As a consequence, and in order to
have an accurate transformation of the PMS evolutionary models of Siess
et al. (2000) into the observational plane, we apply a conversion of these models following an approach similar to that of \cite{girardi02}. We construct, thus, observational models (both tracks and isochrones) for
four assumed metallicities ($Z=0.02$, $0.01$, $0.008$, $0.004$) and
several photometric systems (specifically Johnson, Cousins, HST WFPC2,
HST ACS, HST STIS, HST NICMOS, Galex, Str\"{o}mgren, SDSS, Tycho and
2MASS).

\begin{figure*}
\plottwo{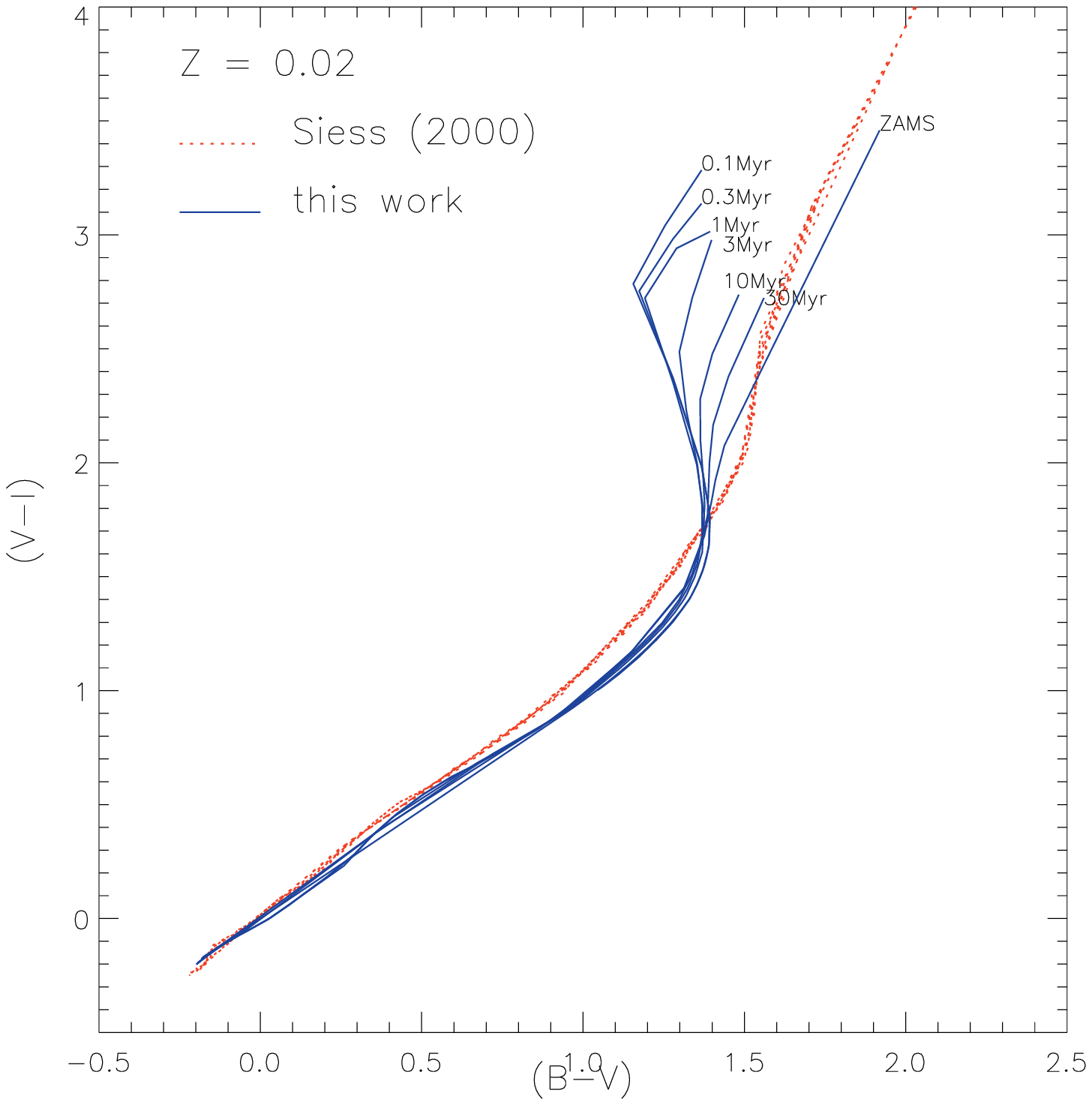}{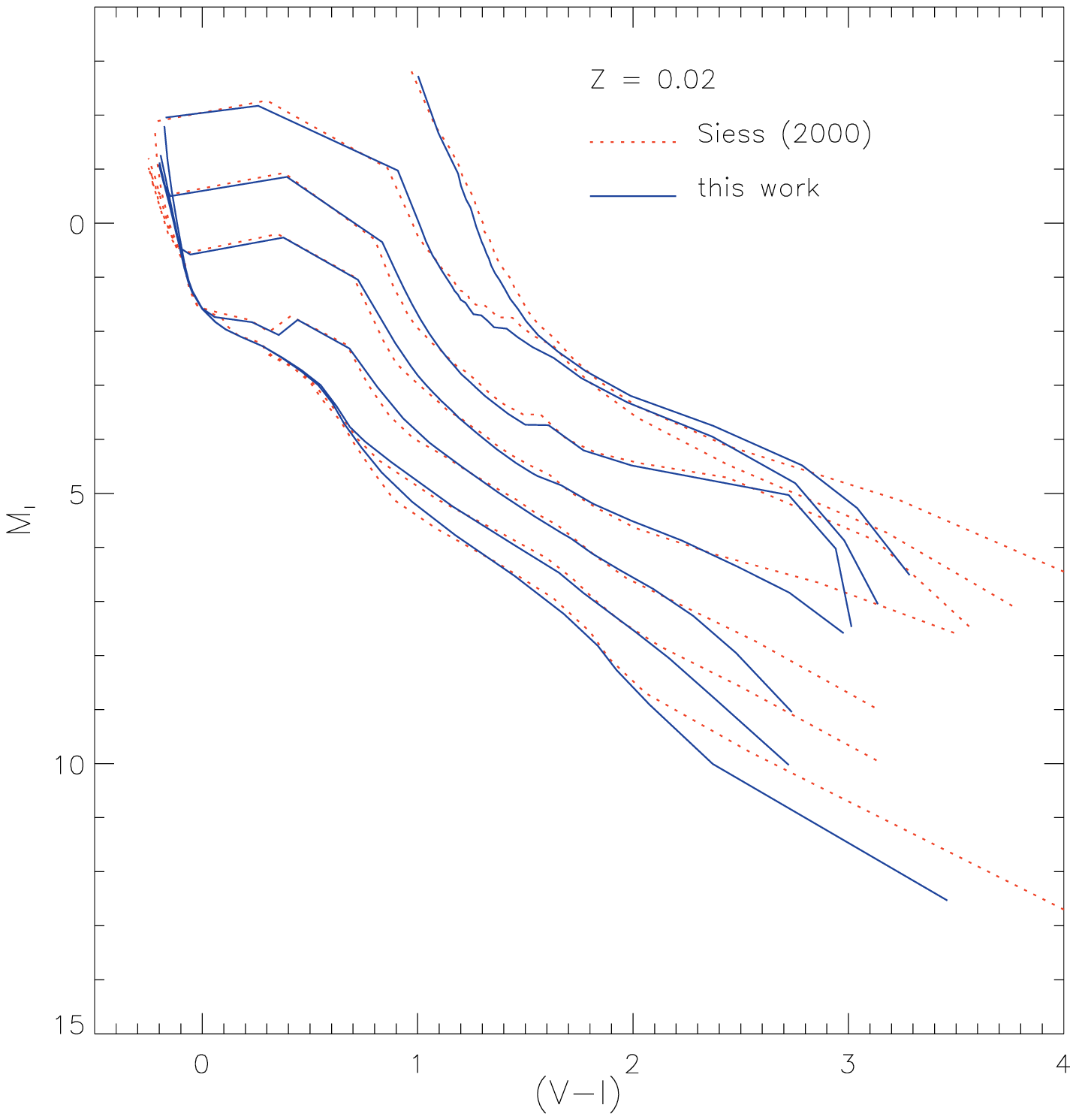}

\caption{Comparison of the evolutionary models for PMS stars developed
in this work with the ones by \citet{siess2000}. {\sl Left}: Color-Color
Diagram of isochrones from the models of \citet{siess2000} for the ZAMS
and ages of $0.1$, $0.3$, $1$, $3$, $10$, $30$ Myr for $Z=0.02$. The
dotted line refers to the original \citet{siess2000} isochrones, in
which a unique temperature $-$ color relation is used, while the solid
lines are the results of our analysis. The differences between the two
treatments are mostly evident at larger colors, at about $T\sim 4000K$,
where M-type optical spectra are dominated by oxide bands, and are
strongly dependent on age due to variations in the surface gravity of
the stars. {\sl Right}: Comparison between the original isochrones of
\citet{siess2000} for $Z=0.02$ and our treatment in the $(V-I)$, $I$
diagrams. For medium masses (top left region) the differences are
negligible, while for low-mass, cold stars the shifts are important. In
particular it is evident that our transformations extend to larger
colors than those of \citet{siess2000}. \label{fig:modcomp}}

\end{figure*}

\subsection{The Method}\label{s:pmstracksmethod}

We apply our conversion method first by collecting a grid of atmospheric
models in the $T_{\rm eff}$, $\log{(g)}$, $[M/H]$ space. These models are
shown in Fig. \ref{fig:spectra}. We selected spectra from
\textsc{NextGen} \citep{nextgen} for low- and medium-temperature stars.
The selected grid covers temperatures 2600~K~$\leq T_{\rm eff}
\leq$~8200~K in steps of 200~K and surface gravities $2.0\leq \log{(g)}
\leq 5.5$ in cgs units, spaced in 0.5 dex. For higher temperatures we used
the grid of models by \cite{kurucz79, kurucz93}, with gravities
$\log{(g)}$ between $3.0$ and $5.0$. This grid covers entirely the range
of stellar parameters in the evolutionary models of Siess et al. (2000).
The considered metallicities cover the values $[M/H]=0.0,\ -0.5,\ -1.0,
-1.5$, representative of stellar populations in the Magellanic Clouds, as
well as in many regions of the Galaxy.  We re-sampled the \textsc{NextGen}
models, the resolution of which is much higher than required for our
purposes of broad-band synthetic photometry, to the wavelength values of
the Kurucz models, and we normalized them to the same units
(erg~cm$^{-2}$~s$^{-1}$~\AA$^{-1}$).

The absolute magnitude in a photometric band $S_{\lambda}$ of a star
with a spectral energy distribution $F_{\lambda}$, and stellar radius
$R$ is given by \begin{equation} \label{equation:first}
M_{S_{\lambda}}=-2.5\log\bigg[\bigg(\frac{R}{10\textrm{pc}}\bigg)^{2}\frac{\displaystyle \int_{
\lambda } \lambda F_{\lambda} S_{ \lambda } 10^{-0.4A_{\lambda}}
\textrm{d} \lambda }{\displaystyle \int_{ \lambda } \lambda f^{0}_{\lambda} S_{
\lambda } \textrm{d} \lambda }\bigg]+ZP_{S_{\lambda}} \end{equation}
where $f^{0}_{\lambda}$ is a reference spectrum that gives a known
apparent magnitude $ZP_{S_{\lambda}}$; in the \textsc{Vegamag}
photometric system, which uses the flux of $\alpha$~Lyr as reference,
$f^{0}_{\lambda}=F_{\lambda , {\rm VEGA}}$ and the zero-points $ZP$ are
close to zero\footnote{Although the historical Vegamag photometric
system was supposed to have zero-points defined so that to impose the
apparent magnitude of Vega equal to zero, improvements in the modeling
of the spectrum of Vega, in its measure and in the definition of
standard throughputs, have introduced small corrections.}.

We assume an extinction $A_{\lambda}=0$, since we are interested in
constructing a general observable grid of evolutionary models for
unreddened objects. Eq.~(\ref{equation:first}) can be written then as:
\begin{eqnarray} \label{equation:second} M_{S_{\lambda}} = &
-5\log\bigg( \frac{\displaystyle R_{\odot} }{\displaystyle 10 {\rm pc}
}\bigg) - 5\log\bigg(\frac{\displaystyle R_{\star}}{\displaystyle
R_{\odot}}\bigg) + B(F_{\lambda},S_{\lambda}) \nonumber \\ = & 43.2337 -
5\log\bigg(\frac{\displaystyle R_{\star}}{\displaystyle R_{\odot}}\bigg)
+ B(F_{\lambda},S_{\lambda}) \end{eqnarray} where \begin{equation}
\label{equation:third}
B(F_{\lambda},S_{\lambda})=-2.5\log\bigg(\frac{\displaystyle \int_{
\lambda } \lambda F_{\lambda} S_{ \lambda } \textrm{d} \lambda
}{\displaystyle \int_{ \lambda } \lambda F_{\lambda , VEGA} S_{ \lambda
} \textrm{d} \lambda}\bigg)+ZP_{S_{\lambda}} \end{equation} The latter
term can be directly calculated having the synthetic spectrum of any
considered star, a calibrated Vega spectrum, the band profile of the
photometric filter considered and the eventual zero-point $ZP$. For the
computation of $B$ we used the \textsc{Gensynphot} code included in the
\textsc{Chorizos} package \citep{chorizos}. Given a set of stellar
parameters ($T_{\rm eff}$, $\log{(g)}$, and $[M/H]$) for a sample of
points (stars) on an evolutionary track or an isochrone, the
corresponding spectra are computed by interpolation on the Cartesian
grid of synthetic models of stellar atmospheres. Then, the integral at
the numerator of Eq.~(\ref{equation:third}) is computed. The final
normalization is performed using a recent reference spectrum of Vega
\citep{bohlinvega} with the corrected zero-points as reported by
\cite{chorizoszp}.

Following this method we converted all the original evolutionary models
of \citet{siess2000} into absolute magnitudes, for the metallicities of
$Z=0.02,\ 0.01,\ 0.008$ and $0.004$ corresponding to $[M/H] \simeq 0.0,
-0.3,\ -0.4,\ -0.7$ respectively. Considering that the original Siess et
al.  models were computed only for $Z=0.01$ and 0.02, we extrapolated
the theoretical isochrones computed for $Z=0.01$ to lower metallicities
with the use of the atmosphere models computed for such metallicities.
Naturally, a more detailed treatment of low-metallicity tracks and
isochrones requires both evolutionary and atmospheric models to be
available at the same considered metallicity. However, for the present
analysis in the LMC, for which $Z\sim 0.008$, the uncertainties are
quite low given the small difference between the metallicity of the
theoretical isochrones (0.01) and that of the atmosphere models (0.008).

\subsection{Comparison with previous conversions}

With our construction of the observational plane for PMS evolutionary
models, we verified that a
single temperature-color relation is not sufficient for stars with
different physical parameters, especially surface gravity.  In
Figure~\ref{fig:modcomp} (left) we present a comparison between the
$(B-V)$, $(V-I)$ color-color diagrams for different isochrones, as they
were derived from the method of \citet{siess2000} (red dotted lines) and
from our method with the use of atmosphere models (blue continuous
lines). In this figure it is shown that the isochrones of
\citet{siess2000} are distributed along one single sequence, due to the
use of monolithic temperature-color relations, which are independent of
the surface gravity. However, our observational isochrones show a spread
for high color terms (low $T_{\rm eff}$), because differences in ages
imply differences in the surface gravities, which produce variations in
the integrated colors according to the atmosphere models we used. The
major differences arise for M-type stars, in which the optical spectra
are dominated by strong oxide absorption bands. Figure~\ref{fig:modcomp}
(right panel) shows these differences on the CMD. As shown in this
figure, the differences between our method and that of \citet{siess2000}
are mostly evident at the red end of the models, where our computations
predict redder colors, and for the youngest models, where the stellar
surface gravities are lower than those for main sequence stars. In
Figure~\ref{fig:modcomp} the example of $Z=0.02$ is shown.

\section{The Initial Mass Function}\label{section:MF}

The mass function (MF) is defined as the number distribution of stars as
a function of mass. While a general observational approach measures the
so called {\sl Present Day Mass Function} (PDMF), a cornerstone for
understanding how stars form is the {\sl Initial Mass Function} (IMF),
which is the mass distribution according to which stars are born. For a
pre-main sequence population younger than the time required by the most
massive stars to disperse it, or to evolve into post-MS evolutionary
phases, the observed mass function coincides with the IMF.

Generally, the IMF is parameterized as follows: \begin{equation}
\xi(M)~\textrm{d}M\propto M^{-(1+x)} \label{equation:imfdef}
\end{equation} \noindent namely, approximating it with a power-law
\citep[e.g.][]{salpeter} or with a series of power-laws, with exponents
changing in different mass ranges \citep{scalo,kroupascience}. In this
section we describe the ``average'' IMF derived for the stellar
populations of the {\sl central region} of LH~95, as it has been defined
in \S~\ref{section:centralregion}.

\begin{figure*}
\epsscale{1.15}
\plotone{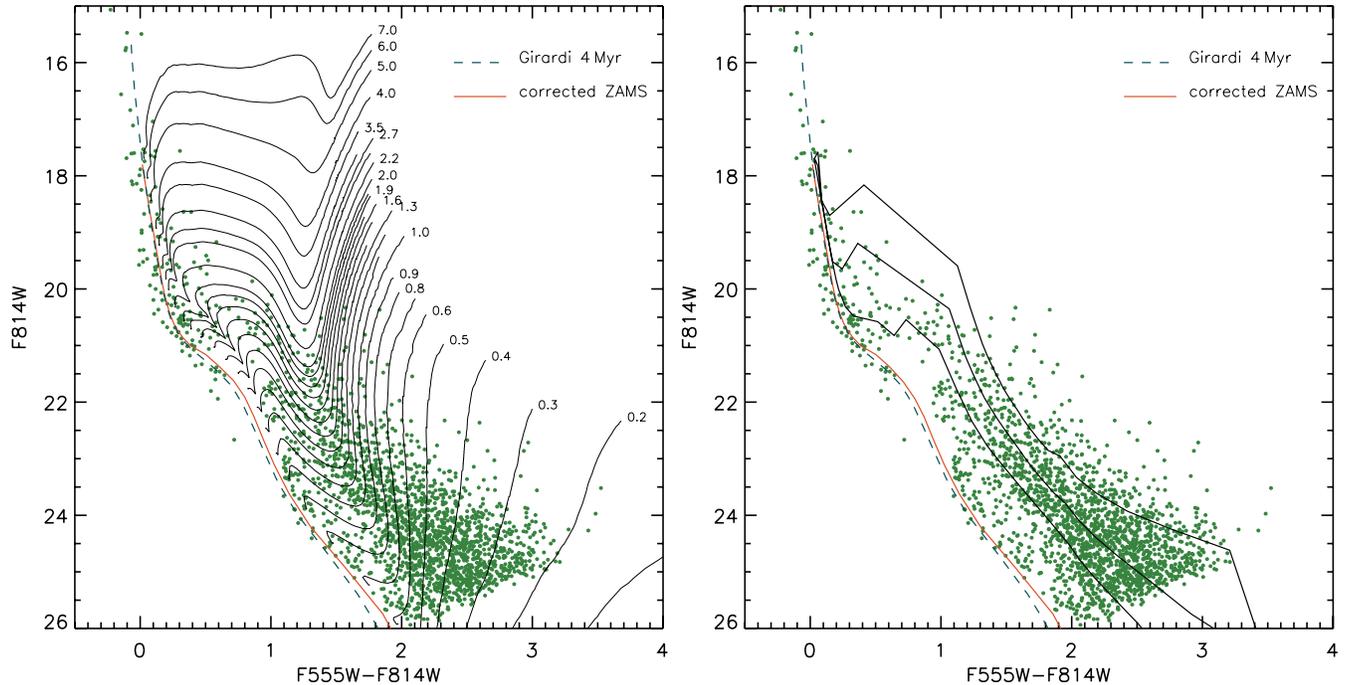}
\caption{The CMD of the cluster area of LH~95 with PMS evolutionary
tracks (left) and isochrones (right) overlayed for a distance modulus
$m-M=18.41$~mag and a mean reddening of $E(F555W-F814W)=0.275$~mag (see
\S~\ref{section:reddening}). The PMS tracks and the corresponding PMS
isochrones and ZAMS are computed by us for the {\sl ACS} filter system,
and for the metallicity of the LMC (\S~\ref{s:pmstracks}). The 4~Myr
isochrone for evolved MS stars is from the Padova grid of evolutionary
models \citep{girardi02}. The comparison of this isochrone with our ZAMS
demonstrates that it practically corresponds to the youngest MS
populations. The selected PMS isochrones overlayed in the right panel have ages of 1.5, 4 and 10 Myr. The central, 4Myr,
corresponds to the the mean age of the observed PMS stars, where the other two delimit the
1$\sigma$ width of the broadening of these stars in the CMD.
\label{fig:CMDwithtracks}}
\end{figure*}

The $F555W-F814W$, $F814W$ CMD of Figure~\ref{fig:CMD-sub} (right panel)
is shown again in Figure~\ref{fig:CMDwithtracks}, with our observable
PMS evolutionary tracks (left panel) and three indicative PMS isochrones
(right panel), computed in \S~\ref{s:pmstracks}, overlayed. In the right
panel of the figure the ZAMS from our grid of models is also plotted
along with the youngest MS isochrone from the Padova grid of
evolutionary models \citep{girardi02} corresponding to $\log{\tau}
\simeq 6.6$. Both isochrones fit to each other very well demonstrating
that the bright MS stars of LH 95 have ages of \lsim~4~Myr. As far as
the PMS stars of LH~95 concerns, their positions in the CMD corresponds
to an age-spread that covers ages between $\sim$~1.5 and 10~Myr
according to our PMS models. A simple statistics on the distribution of
the ages of these PMS stars, based on their CMD positions, results to a
normal distribution of ages which peaks at about 4~Myr. However, if this
age-spread is true or not can be the subject of debate, as we discuss in
\S~\ref{section:fieldsubtractio2}. Therefore, based on the PMS
population of LH 95 we define an indicative age for LH~95 of $\log{\tau}
= 6.6 \pm 0.4$. Corresponding PMS isochrones for ages of 1.5, 4 and
10~Myr from our grid of observable PMS models are overlayed on the CMD
in the right panel of Figure~\ref{fig:CMDwithtracks}. A comparison of
this age, which also represents the individual subclusters of LH~95,
with the corresponding dynamical crossing time, estimated in
\S~\ref{section:centralregion}, shows that all sub-clusters, as well as
the whole central region have an age comparable to their crossing times,
indicating that these systems are under `mixing' process if not already
`mixed' and therefore we probably do not observe them close to their
initial morphology.

\subsection{Construction of the IMF}\label{section:IMFconstruction}

Based on our new set of PMS models computed above for the ACS
photometric bands and the average LMC metallicity ($Z\sim 0.008$), we
assigned a value of mass to all the stars identified in the region of
LH~95, by comparing our observed data with the evolutionary tracks of
these models. For this comparison we assumed a distance modulus of
$m-M=18.41$~mag, and an average reddening of $E(F555W-F814W)=0.275$~mag,
as it is derived in \S~\ref{section:reddening}.  According to Figure \ref{fig:CMDwithtracks},
our photometry reveals PMS stars with masses as low as 0.2~M$_{\odot}$.
Our sample, though, as we discuss later, is complete for masses down to
$\simeq$~0.43~M{\solar}.
For the PMS stars we assign a mass measurement to each observed star
according to their positions in the CMD by interpolating between the PMS
evolutionary tracks shown in Figure~\ref{fig:CMDwithtracks}.

The position of the brightest main sequence stars is not covered by our
PMS evolutionary models, limited for masses below 7M$_{\odot}$.
Therefore, we derive the masses of these stars directly from the Padova
4~Myr isochrone -- already used for the determination of the average
reddening in \S~\ref{section:reddening} -- shifted in the direction of
the reddening vector to reach individually each one of these stars. The
good agreement between this Padova isochrone and the corrected ZAMS from
the Siess models is evident from Figure \ref{fig:CMDwithtracks}. While
we de-redden the most massive MS stars according to their individual
position in the CMD, their mass estimation is based solely on
photometry. A more accurate estimation of the mass for massive stars
certainly requires spectroscopy, because for such hot stars the
broad-band colors are degenerate \citep{massey95}, and naturally this
effect may influence the mass estimation for the brightest MS stars in
our sample and the corresponding IMF. We discuss the influence of the
use of photometry alone to the slope of the derived IMF in
\S\ref{section:IMFfitting}.

With the masses of both the MS and the PMS stars in our
sample available the construction of their IMF is in principle
straightforward, by binning the stars according to their masses and
fitting the derived distribution. However, there are considerations,
which affect the construction and interpretation of the IMF, and which
should be taken into account. We discuss these issues below.

\begin{itemize}

\item[{(i)}] An important issue to be considered is {\sl the
decontamination process of the stellar sample} in the region of the
association from the field populations. The Monte Carlo method for the
field subtraction, as applied in \S~\ref{section:fieldsubtractio2}, is
based on probabilities and therefore it is a stochastic process. As a
consequence, the number of stellar members of the system, which are
counted in an arbitrary mass bin is not determined in an unambiguous
manner. Moreover, the statistical significance of this number is not
determined by the fraction of stars considered as system members over the
total number of stars,  but it depends also on the probability that {\sl
all} the  observed stars in the bin are system members. This dependence
affects specifically the MS populations with 19~\lsim~$m_{\rm
F814W}$/mag~\lsim~22, which as it is shown in Figure~\ref{fig:CMD-sub} are
well mixed with field populations, implying a higher uncertainty in the
determination of the fraction of system members.
In the faint part of the CMD, system and field populations are in
general well distinguished, and therefore the uncertainties in the
numbers of PMS stars that are true system-members are insignificant.
However, there is an important possibility that a fraction of the binary
sequence, evident in the CMD for the LMC field population (see
Figures~\ref{fig:CMD-sub} and \ref{fig:CMDwithtracks}), contaminates the
selected sample of PMS star-members of the system. We consider this bias
and report on its effect in \S~\ref{section:IMFfitting}

\item[{(ii)}] {\sl The fitting process of the IMF}. Typical linear
regression methods, which are used for obtaining a functional form of an
observed IMF, such as the $\chi^2$-method, consider the presence of
measurement errors and are based on the assumption that the uncertainty
associated to each point follows a Gaussian distribution. However, in
this case the uncertainty in the number of counts within a mass bin is
the overall effect of both the Poissonian error that naturally comes from the
counting process, and the uncertainty that arises from the field
subtraction, as discussed above. Although the first can be well
approximated with a Gaussian for large $n$, the latter depends on the
positions of stars and tracks in the CMDs. Moreover, in the high mass
regime, where the number statistics is poor, the Poissonian error cannot
be treated as Gaussian, requiring the asymmetry of the distribution to
be taken into account.

\item[{(iii)}] {\sl The completeness-correction process}. In the
low-mass regime corrections for the incompleteness of the stellar sample
are required to estimate the actual number of stars. However, given the
width of the low-mass evolutionary tracks in the CMD shown in
Figure~\ref{fig:CMDwithtracks} and the 2-dimensional variation of the
completeness itself in the CMD (in both magnitudes and colors; see
\S~\ref{section:completeness}), this correction is not unique for all
the stars counted in a considered mass bin.

\end{itemize}

We construct the system IMF of the association LH~95, with a special care in
addressing the above issues, as follows. We first determine a grid of
logarithmic mass bins, for which we choose variable width in $\log(M)$,
that increases from $0.04$~dex at low masses to $0.25$~dex for the
higher masses. The counting of the stars is made in variable-sized bins
because such bins yield very small biases, which are only weakly
dependent on the number of stars, in contrast to uniformly binned data
\citep{imfbias}. For each $i^{\rm th}$ bin a number of stars $N_i$ is
derived. In order to achieve a statistically correct sampling of points
for the constructed IMF, we simulate for each bin the probability
distribution expected for every $N_i$ with a sample of 1000 points
randomly selected from a Poisson distribution with mean $N_i$.

We then consider the position in the CMD of all the stars in each bin,
and we compute for each star its completeness $C$ (the ratio between the number of sources detected and the total number of sources) as a function of both
magnitude {\sl and} color according to the completeness measurements
described in \S~\ref{section:completeness}, for the central region of LH~95.
Let us consider the $j^{\rm th}$ star counted in the
$i^{\rm th}$ bin. The star is located in a specific position of the CMD,
where completeness is $C_j$, meaning that the total number
number of stars corresponding to $j$ is simply $1/C_j$. Consequently,
for the total number of stars counted in the bin, $N_i$, the
completeness correction is:
\begin{equation} F_i=\frac{\displaystyle
\sum_{j}^{N_i}\frac{\displaystyle 1}{\displaystyle C_j}}{\displaystyle
N_i} \end{equation}
which is a quantity $\geq1$.

We multiply our Poissonian sample associated to each $i^{\rm th}$ bin
by  $F_i$, and we divide by the width of the bin in order to express
numbers in units of number of stars per solar mass\footnote{It is
worth-noting that the completeness correction should be applied {\sl
after} simulating the Poisson distribution, and not before. For example,
if one star of $25\%$ completeness is counted into the $i^{\rm th}$ bin,
the true number of sources associated to the $N_i$ sample is $4$, with
an uncertainty of $\sqrt{1}*4=4$ and not $\sqrt{4}=2$.}. In order to
take into account the additional error introduced by the stochasticity
of the Monte Carlo field-subtraction
(\S~\ref{section:fieldsubtractio2}), we iterate the technique of field
removal $200$ times, each time repeating the procedure explained in the
previous paragraphs, namely, counting the $N_i$ number of stars in each
$i^{\rm th}$ bin, sampling a Poisson random population associated to it,
and computing and applying the completeness correction.

The outcome of this approach is that for each mass-bin, instead of a
single point-error combination, we have a population of 200,000 random
IMF points, derived according to the true statistical distribution of
the expected number of stars in the bin. Clearly, the mean and the
standard deviation of each of these distributions represent the IMF data
points and error bars. The main advantage of this technique is that we
override the non-Gaussian behavior of the mass-bin error bars. Specifically,
now for every mass-bin we do not have one point associated with an error
that cannot be treated by the regression theory, but a numerous set of
points with no uncertainties associated to them, and consequently with
arbitrary equal weights for all mass-bins. Naturally, this allows a more
accurate fitting of the mass function.

\begin{figure}
\epsscale{1.15}
\plotone{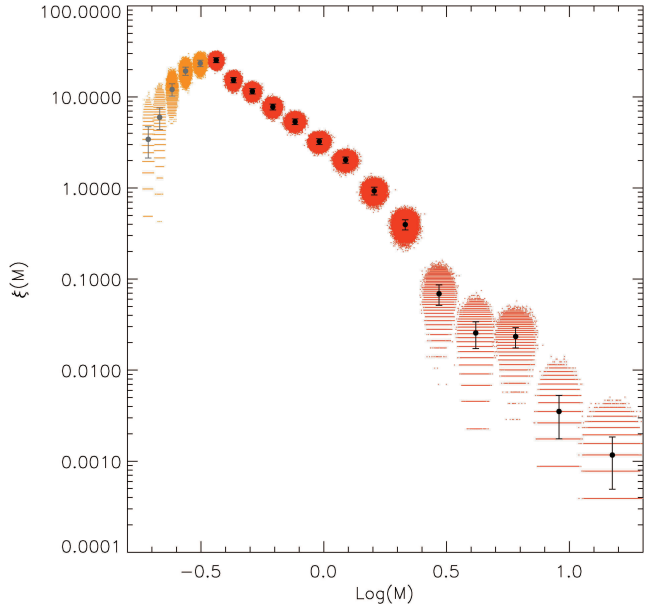}
\caption{The system IMF plot of LH~95, which demonstrates our method for
treating the distribution of the expected number of stars per mass-bin
according to uncertainties due to counting, incompleteness and field
subtraction. Each mass-bin is represented by the distribution along the
$y-$axis of a set of 200,000 randomly selected points, which map the
true statistical uncertainties of star counts. An additional Gaussian
spread with $\sigma = 1/8$ of the bin width along the $x-$axis has been
artificially added for each bin to better highlight the behavior of the
individual distributions. The asymmetry of the statistical distributions
especially for the high-mass end of the IMF is demonstrated by the
drop-like shape of the distributions. The very-low mass end (light-gray
points) appears to drop due to the limitation of our completeness
correction technique in CMD regions where no stars are detected. Units
of the IMF are number of stars per solar mass per pc$^2$
\label{fig:MFmontecarlo}.}
\end{figure}

In Figure~\ref{fig:MFmontecarlo} we visualize the concept of this
approach. For every mass-bin the set of random points, representing the
distribution of the expected $\xi(M)$ is drawn with tiny colored dots.
In this plot an artificial spread along the $x$-axis (a random Gaussian
shift with $\sigma = 1/8$ of the bin width) is introduced to better
highlight the distributions along the $y-$axis. The black dots with
error bars show the average and standard deviation of each sample of
points. The derived numbers are reliable for $M$~\gsim~0.35~M$_\odot$,
whereas for the lower masses (first five mass bins from the left) the completeness corrections
are not sufficient. This limitation is evident from the CMD of
Figure~\ref{fig:CMDwithtracks}, where it can be seen that for the
evolutionary tracks for very low masses ($M$~\lsim~0.35~M$_\odot$), the
completeness is extremely low, so that no stars can be detected, limiting
the ability to recover the true number distributions for such masses.

\subsection{IMF model fitting}
\label{section:IMFfitting}

A commonly used procedure to derive the functional form of an observed
IMF is to fit a predefined function to the individual numbers of stars
originally counted per mass-bin taking into account the corresponding
counting errors. However, in our case, we have simulated the statistical
distribution of $\xi(M)$ per mass-bin taking into account the additional
errors due to incompleteness and field subtraction. Within our statistical approach we perform 200 repetitions of the Monte
Carlo field subtraction method and for each of them we map the
Poissonian statistics 1000 times. The outcome is a non-analytical probability
distribution for each mass-bin that represents the true expected
statistical uncertainties due to the issues discussed in \S\ref{section:IMFconstruction}. This process
provided a large sample of 200,000 points per mass-bin, and naturally
the best-fitting function can be derived more accurately with the use of
these data, rather than the original 14 bins. The only assumption that
we make is that the functional form of the IMF of LH~95 can be correctly
represented by power-laws, according to the parameterization given in
Eq.~(\ref{equation:imfdef}), and therefore we search for the best
multiple power-law which may reproduce our data. As seen in
Figure~\ref{fig:MFmontecarlo} the IMF seems to be more shallow for the low
masses ($M\lesssim1$~M$_\odot$) than for intermediate and high masses,
implying that it could be reproduced by a two-phase power-law. We
investigate, however, whether a two-phase power-law is sufficient or a
three-phase power-law, such as that proposed by \cite{scalo98}, is
required to represent our IMF.

In order to find the best fit of two- and three-phase power-laws to our
IMF we applied a Levenberg-Marquard non-linear least square minimization
technique \citep{Levenberg44, Marquardt63}. For this fit we consider
only the mass-bins of the IMF for which the completeness correction
allowed us to reconstruct the actual number of stars, meaning for
$\log{(M/{\rm M}_\odot)}>-0.5$. The slopes of the power-laws, as well as
the position of the break points along the abscissa are the free
parameters of our fit algorithm. All 200,000 points of each
mass-bin are considered of equal weight. However, since a two-phase
power-law is a sub-case of the three-phase power-law, the sum-of-squares
of the latter is always lower. Therefore, the results of the fit
algorithm for each case are normalized in a rigorous manner by
performing the so-called statistical ``F test''.

This test compares the relative increase in the residual sum of squares
by reducing the complexity of the model with the relative increase of
degrees of freedom, namely, the differences between the number of data
points and the free parameters. Factor $F$ is, thus, computed as
\begin{equation} \label{equation:Ftest} F=\frac{\textstyle
\frac{\textstyle RSS_2-RSS_3}{\textstyle
p_3-p_2}}{\textstyle\frac{\textstyle RSS_3}{\textstyle n-p_3}}
\end{equation} \noindent where $RSS_2$ and $RSS_3$ are the residual sum
of squares for the two- and three-phase power-law best fit models,
$p_2=4$ and $p_3=6$ are the corresponding number of free parameters, and
$n$ the total number of points, which, in our case, is the effective
number of considered mass bins (14) and not the total number of
variables used in the fit (200,000~$\times$~14), given that the
additional multiplicity has been introduced to map the non Gaussian
errors. Typically, if the simpler model is valid (which means that there
is no need to add parameters), then $F\simeq1$, whereas if the complex
model represents the distribution better, then $F\gg1$. We obtain
$F=1.15$. The probability, $P$, that the null hypothesis, which is ``the
improvement found adding additional parameters is solely due to chance''
(or ``the simpler model is good enough'') should be rejected is given by
the value of the cumulative {\sl Snedecor's} $F$ distribution with
$(p_3-p_3,n-p_3)$ degrees of freedom. In our case it is $P=0.65$, while,
generally, a value of $P>0.95$ is required to state with sufficient
significance, that the complex model is better. Under these
circumstances, we conclude that our system IMF is best approximated by a
two-phase power-law.

\begin{figure}
\epsscale{1.15}
\plotone{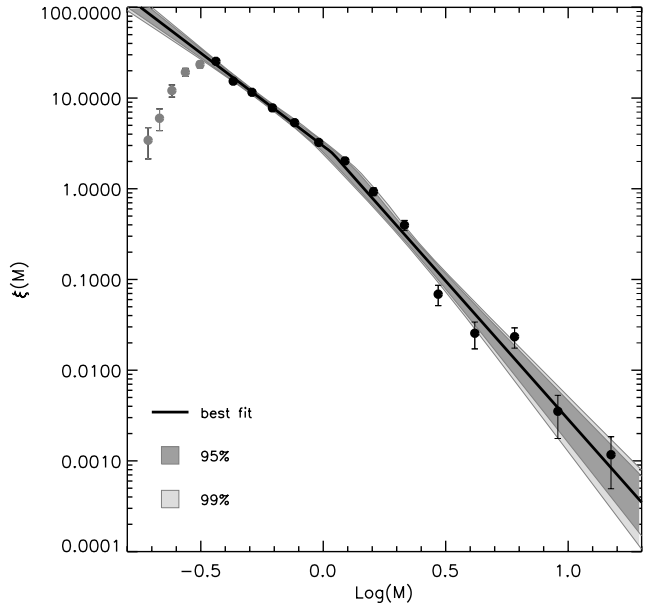}
\caption{ The system IMF of the association LH~95. The best two-phase power-law
fit derived in \S~\ref{section:IMFfitting} is drawn with a solid line,
while the shaded areas represent the $95\%$ and $99\%$ confidence
uncertainties in the slope determination and the break point. Units of
the IMF are logarithmic number of stars per solar mass per pc$^2$. The
corresponding IMF slopes and their uncertainties are given in
Table~\ref{t:IMF_expression}.
\label{fig:MF}}
\end{figure}

We compute the uncertainty associated to this result, by utilizing a
sampling technique. For each mass-bin we pick randomly one out of the
200,000 points from the corresponding distribution of $\xi(M)$ in that
bin, and we perform a two-phase power-law fit as described above,
obtaining the two measurements for the slope, and the point ($\log{M}$,
$\log{\xi(M)}$) where the change of the slope occurs. After repeating
this process for a substantial amount of times, we obtain a
statistically significant sample for the values of the two slopes and
the point of change, and consequently the relative confidence level of
our fit. Figure~\ref{fig:MF} shows the result of our fit to the system IMF of
LH~95. The best two-phase power-law fit is shown as a solid line, while
the uncertainty of this fit is demonstrated by the grey shaded areas for
confidence of 95\% (dark grey) and 99\% (light grey) respectively. In
Table~\ref{t:IMF_expression} we give the system IMF slopes $x$ and their
uncertainties with the corresponding mass ranges. The 1$\sigma$ error in
the determination of the break point is 0.07 dex.

Another bias that should be considered in the construction of the system
IMF of LH~95 is the possible contamination of our selected sample of PMS
stars by a binary sequence of the LMC field MS population, discussed in
\S~\ref{section:IMFconstruction}. In order to quantify the effect of
this bias we select this part of the CMD and remove the corresponding
stars. Specifically, we remove all stars bluer than the borders
defined by the points (F555W$-$F814W, F814W)~$=$~(0.9, 22.9), (1.65, 24.3) and (2.2,
26.9) on the CMD. The number of these stars changes according to the
stochastic Monte Carlo field subtraction; on average they count to 82
members, less than 4\% of the selected PMS population. Then, we reapply
the analysis described in \S\S~\ref{section:IMFconstruction} and
\ref{section:IMFfitting} for the construction and model-fitting of the
system IMF. We find that this new IMF does not differ significantly from
the one shown in Figure~\ref{fig:MF}, and its shape is not different
from that described with the slopes of Table~\ref{t:IMF_expression}.
Specifically, this IMF has 99\%-confidence slopes $x \simeq 1.16$ for
$-0.5<\log(M/{\rm M}_\odot)<0.04$, and $x \simeq 2.03$ for $\log(M/{\rm
M}_\odot)>0.04$, indistinguishable from the slopes of
Table~\ref{t:IMF_expression}. Our analysis on the sub-clusters of LH~95
showed that the contamination of our sample by a MS binary sequence does
not alter also the shape of their individual system IMFs, with slopes as well
indistinguishable from those given in Table
\ref{t:IMF_expression_subcluster}.

The lack of spectroscopy and the use of broad-band photometry alone for
the mass estimation of massive MS stars is found to lead to steeper IMF
slopes \citep{massey95}, and this may affect the overall intermediate-
and high-mass slope of the IMF of LH~95. However, considering that in
their analysis \cite{massey95} show that this effect is significant for
stars more massive than 15~M{\solar}, the only mass-bin in our IMF
that can be biased is the most massive one, comprising stars with
$11$~\lsim~$M$/M{\solar}~\lsim~$20$. In addition, the IMF derived from
photometry alone for high-mass stars in star-forming regions of the
Magellanic Clouds {\sl is found to be also quite shallow and not steep},
and within the observed IMF slope variations in such regions (see
discussion in \S\ref{ssec:imf-z}). Under these circumstances, the
corresponding uncertainty in our derived system IMF slope due to the use of
photometry alone should not be expected to be larger than those given in
Table~3.

\begin{deluxetable}{rrrr}
\tablewidth{0pt}
\tablecaption{The slope of the system IMF of the association
LH~95\label{t:IMF_expression}}
\tablehead{
\colhead{mass range}&
\colhead{$x$}&
\colhead{$\sigma_{95\%}$}&
\colhead{$\sigma_{99\%}$}
}
\startdata
$-0.5<\log(M/{\rm M}_\odot)<0.04$ & 1.05 & $^{-0.20}_{+0.15}$ &
$^{-0.29}_{+0.20}$ \\
\\
$\log(M/{\rm M}_\odot)>0.04$ & 2.05 & $^{-0.28}_{+0.39}$ &
$^{-0.32}_{+0.53}$ \\
\enddata
\tablecomments{A Salpeter (1955) IMF would have a slope of $x=1.35$.}
\end{deluxetable}

\section{Discussion}
\label{section:IMFdiscussion}

\begin{figure*}
\epsscale{1.15}
\plotone{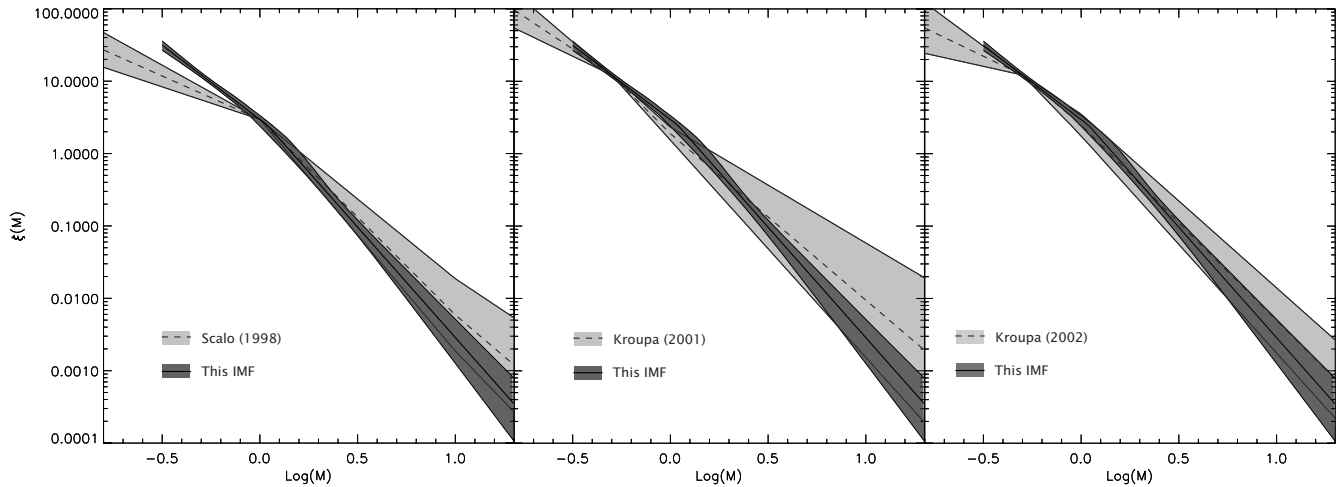}
\caption{Comparison between the derived LH~95 system IMF and the average IMF of
\cite{scalo98} (left), \cite{kroupa} (middle), as well as the universal
IMF of \citet{kroupascience} (right).\label{fig:MFwithkroupa}}
\end{figure*}

The IMF constructed in the previous section is the most complete
extragalactic system IMF down to the sub-solar regime derived so far. Specifically, our
correction for incompleteness in the derivation of $\xi(M)$ actually
provides a reliable IMF down to $M\simeq 0.43M_{\odot}$, as discussed in
\S\ref{section:IMFconstruction}. Probably, the most interesting result
concerning this IMF is the change in its slope, which we verified
statistically for stars in the the subsolar mass range. The IMF slope,
$x$, decreases for these stars by one unit.  An analogous behavior is
generally found in related studies in the Galaxy. Comprehensive studies
which summarize the average properties of the galactic IMF are those of
\citet{scalo}, revisited in \cite{scalo98} and
\citet[2001,][]{kroupascience}.  The latter investigations are rather complete,
since they cover stellar mass distributions in a wide mass range, from
brown dwarfs with $M\sim0.01M_\odot$ to the most massive stars.
According to \citet{kroupascience}, the IMF slope $x$ (Eq.
\ref{equation:imfdef}) changes from $x=-0.7$ in the sub-stellar mass range, to
$x=0.3$ for masses between 0.08~M$_\odot$ and 0.5~M\solar, $x=1.3$ for
0.5~M\solar~$<M<$~1~M\solar, and $x=1.7$ for stars of higher masses.
This IMF is generally characterized as the Galactic average, in the
sense that it is reasonably valid for different regions of the Galaxy.
A comparison between the system IMF we construct here for LH~95 and that of
\cite{scalo98}, as well as the universal IMF of \cite{kroupa} and the
average Galactic IMF of \citet{kroupascience}, is shown in
Figure~\ref{fig:MFwithkroupa}. We scaled all other IMFs to match our mass
distribution at $\log(M/{\rm M}_\odot)=-0.3$. In
Figure~\ref{fig:MFwithkroupa} an overall agreement between the various IMFs is quite evident.

The mass limit of the breaking point in our system IMF ($\sim$~1~M{\solar})
coincides with the ``standard Galactic field IMF'', which has been
reported to have a change of slope at $1 M\solar$ \citep{kroupascience},
in agreement also with the knee of the IMF derived by \citet{scalo98}.
In addition, there is a second knee in the Kroupa IMF, which occurs at
the lower mass-limit of 0.5~M{\solar}. Since we cannot constrain the system IMF
below $0.43M\solar$ we do not identify this second knee. The slope of our
IMF for stars with $M$~\gsim~1~M{\solar} falls within the range of
slopes found by both Scalo and Kroupa.
This is in line with the recent finding that the system IMF is
essentially equal to the stellar IMF for intermediate and massive stars
\citep{weidner08}. Our system IMF, however,
is somewhat steeper, as an indication of a bottom-heavy IMF, but the use
of photometry alone for the estimation of the masses for the brightest
stars ($M$~\gsim~15~M{\solar}) may bias the slope to steeper values (see
\S\ref{section:IMFfitting}). As far as the slope of our IMF in the
sub-solar regime is concerned, it is found slightly more shallow than
that of these authors.
Specifically, our IMF is by $\Delta x = 0.1$~-~$0.3$ more shallow than
the stellar IMF \citep[as seen from the theoretical $\alpha$-plots
of][]{kroupa}, making the resulting stellar IMF for LH~95
indistinguishable to the Galactic-field stellar IMF
\citep[e.g.][]{kroupascience}.
It should be noted that both Scalo and Kroupa
IMFs extend down to masses as low as the hydrogen burning limit
($M\simeq$~0.08~M$_\odot$), well below our detection limit. In general,
one can conclude that the system IMF we derive for LH~95 is in agreement with
the average galactic IMF in the entire mass range we studied, from
0.43~M$_\odot$ to $\sim$~20~M$_\odot$.

The high-mass IMF of LH~95 was constructed earlier from direct CCD
imaging in $BVR$ with the 1-m telescope at Siding Spring Observatory by
\citet{gouliermis02}. These authors provide the field-subtracted
completeness-corrected IMF of the system for stars with masses down to
$\sim 2$~M{\solar}, and they find that this IMF can be fitted with a
single power-law with slope $x=1.2 \pm 0.3$ for stars with masses up to
about 17~M{\solar}. This result indicates a more shallow slope $x$ than
what we found for intermediate and high mass stars ($x=2.05$). However,
in our analysis we included masses as low as $1M_\odot$, extending the
mass range investigated in \citet{gouliermis02}. Indeed, if we exclude
the mass bins from 1 to 2~M$_\odot$ ($0<\log{(M/{\rm M}_\odot)}<0.3$)
from our analysis we derive a more shallow value for $x$ for the
intermediate- and high-mass stars. The effect of a steeper IMF for the
mass range between 1 and 2 M{\solar} is evident in
Figs.~\ref{fig:MFmontecarlo} and \ref{fig:MF}. After excluding stars of $M<2$~M$_\odot$ from the
considered stellar sample and performing a single
power-law fit on the resulting $\xi(M)$, we find an IMF slope of
$x\simeq1.55$, comparable to that of \citet{gouliermis02}.

Concerning our derived IMF for the central region of LH~95 it is
important to assess the accuracy of our findings, in relation with the
possible sources of uncertainties. One of the most important error
sources normally comes from the incomplete stellar samples. However, in our case we
consider the completeness as it has been derived from artificial star
tests in the central part of LH~95, where the crowding is higher
(Figure~\ref{fig:photomerrors_completeness}b), and we performed rigorous
corrections for incompleteness taking into account variations of
completeness for all stellar positions in the CMD. As far as the
very-low mass stars is concerned, we do not consider those with $M$
\lsim 0.35~M$_\odot$, where our completeness correction is limited.

\subsection{Stellar Binarity}

An important source of uncertainty in the construction of the IMF is
also the presence of unresolved binaries. Unresolved binarity, can modify the shape of the observed mass function
so that it appears more shallow than what it actually is.
For this reason, we stress that the IMF reported in Table \ref{t:IMF_expression} must be considered as the {\sl system} IMF of our region, and not a {\sl stellar} IMF as which one refers when the effect due to binarity is removed.
Clearly it is
not possible to correct this effect without the knowledge of the actual
binary fraction of the studied population, but one can estimate the
overall effect on the final mass distribution. \citet{sagar91} applied
Monte Carlo simulations on artificial stellar populations, in order to
investigate the change in the mass function slope $\Delta x$ due to
binarity for stars in the mass range $2<M/M\solar<14$. These authors found that the result depends on both the binary
fraction $f$ and the original slope $x_0$ of the mass function.
Specifically, for an original slope $x_0=1.5$ and a binary fraction
$f=0.5$ \citet{sagar91} found a slope change $\Delta x\simeq 0.34$,
while for $x_0=0.5$ and $f=0.5$, $\Delta x\simeq 0.5$.
On the other hand, \citet{kroupa91} in their investigation
on the low-mass IMF in the solar neighborhood, showed that binaries
can make a major effect in mass function determinations below 1~M{\solar}. Consequently, the expected changes in the slope due to unresolved binarity may introduce uncertainties in the derived
slopes given in Table~\ref{t:IMF_expression}, but not large enough to justify the intrinsic change of slope that we observe in our IMF for LH~95 at 1~M{\solar}.
In any case, as discussed by \citet{kroupascience}, most of the studied
mass functions in the literature are not corrected for binarity, mostly
due to the high uncertainty in the binary fraction. The Galactic IMF derived by \citep{kroupa,kroupascience}, however, has
been corrected for unresolved multiple systems.
A more recent study of the binary effect on the massive-star IMF by
\citet{weidner08} shows that a power-law IMF is not
significantly affected even by large numbers of unresolved binaries or
higher order multiples.

In order to analyze the effect of unresolved binarity in the measured
slope of the system IMF in the sub-solar regime, we apply a Monte-Carlo
technique adopted from the work of \citet{sagar91} and
\citet{kroupa91,kroupa93}. For this method we assume a pre-defined set
of functions for the IMF, and in order to be consistent with those we
find for the system IMF, as well as with the Galactic IMF of
\citet{kroupascience}, we consider three-phase power laws with break
points at 0.3~M{\solar} and 1~M{\solar}. We limit our mass samples in
the stellar regime with $M >$~0.08~M{\solar}. We let the slopes $x_1$,
$x_2$, $x_3$ within the three selected mass ranges to change
respectively in the range $-1<x_1<1.5$ for $0.08<M/{\rm M}_{\odot}<0.3$,
$0<x_2<2$ for $0.3<M/{\rm M}_{\odot}<1$ and $1<x_3<2$ for $M/{\rm
M}_{\odot}>1$. For each of these mass distributions, we generate a
sample of 50,000 stars positioned in the CMD along our 4~Myr PMS
isochrone (corresponding to the age of the system). We assume binary
fraction, $f$, between 0 and 1, and we randomly pair couples of stars
from the simulated distributions. We then compute the new position in
the CMD of these binary systems and assign new masses as if they were
single stars. The out-coming mass distributions are fitted with a
3-phase power law imposing the same position of the break points.

We isolate the models for which the derived system IMF in the range
$0.3<M/{\rm M}_{\odot}<1$ is compatible with the one measured for LH~95
within the confidence interval of the latter. Naturally there are
multiple solutions, given the lack of knowledge about the actual IMF of
LH~95, its shape at lower masses, and the binarity fraction appropriate
for this system. However, we can constrain the uncertainties due to
unresolved binarity in our measured slope, $x_2$, for the subsolar
regime $0.3<M/{\rm M}_{\odot}<1$. The IMF slope within this mass range
is mainly affected by the value of $f$. If we assume $x_1$ to be that of
the Galactic-field IMF, the slope $x_2$ in this mass range changes by
$\Delta x_2\simeq0.45$ for $f=1$. This change decreases to $\Delta
x_2\simeq0.35$ for $f=0.6$, and $\Delta x_2\simeq0.25$ for $f=0.4$.

A change in the original IMF for masses larger than Solar does not
affect our results by more than $\Delta x_3 = 0.1$, whereas changing the
slope at very low masses ($0.08<M/{\rm M}_{\odot}<0.3$) affects
significantly the slope $x_2$, diminishing it when $x_1$ becomes much
steeper or shallower than the value of \citet{kroupascience} IMF. Even
in this case, the effect is always less than $\Delta x_3 = 0.2$.
Therefore, we conclude that for a reasonable binary fraction of $f
\simeq 0.5$, unresolved binarity can bias our IMF slope, $x$, in the
subsolar regime on average by no more than 0.3 units. As a consequence
unresolved binarity does not affect significantly the measured IMF for
LH~95 within the estimated errors, and thus the corrected stellar IMF
for LH~95 remains indistinguishable from the Galactic IMF.

\subsection{Metallicity effects}\label{ssec:imf-z}

Recent findings suggest that variations in the metallicity of a
star-forming cloud might cause variations in the IMF slope for stars
with masses $M<0.7$~M{\solar}, in the sense that Galactic regions with
higher $[Fe/H]$ appear to produce more low-mass stars
\citep{kroupascience}. This is further supported by the measured IMF in
Galactic open clusters \citep{barrado2001}, globular clusters
\citep{piotto1999} and old and relatively metal-poor thick disk stars
\citep{reyle2001}. \citet{kroupascience} proposes a systematic
metallicity dependence of the IMF exponent of the form $x\approx 0.3 +
\Delta x [Fe/H]$, where $\Delta x \approx 0.5$. If we apply this
dependence to the LMC, for which $[Fe/H] \simeq -0.4$, we find that the
slope of the average IMF in the LMC for masses $M<0.7$~M{\solar} should
be flattened to the value $x\approx 0.1$. The flattening of the IMF of
the LMC general field for stars with $M<0.7$~M{\solar} has been
discussed by \citet{gouliermis06a}, who found only indications of such a
change in the IMF slope, because of observational limitations. Our data,
on the other hand, being deeper, show a constant IMF slope in the
low-mass regime down to the detection limit of $M \simeq 0.43$~M{\solar}
with no indication of flattening for $M<0.7$~M{\solar}. Consequently, we
cannot verify any metallicity dependence of the sub-solar IMF in the
LMC, at least based on the formula of \citet{kroupascience}. Considering
that this result is based on a single stellar system, more studies of
low-mass populations in the Magellanic Clouds may provide a better
constraint on any systematic metallicity-dependence of the low-mass IMF.

As far as the high-mass regime is concerned, while the IMF of LH~95
seems to be steeper than that derived from ground-based observations by
\citet{gouliermis02}, our analysis (\S\ref{section:IMFdiscussion}) shows
that if we consider only stars more massive than $3$~M{\solar} the IMF of
LH~95 has a Salpeter-like slope, comparable to that of
\citet{gouliermis02}. While, as discussed in \S\ref{section:IMFfitting},
the use of photometry alone leads to steeper IMFs, there are several
photometric studies of associations in the LMC, which derive {\sl a
high-mass IMF that is not steeper, but more shallow} than the typical
Salpeter IMF \citep[e.g.][]{hill1994, oey1996, dolphin98}. Taking into
account all available studies for the intermediate- and high-mass stars
with $M$~\gsim~7~M{\solar}, both spectroscopic {\sl and} photometric
\citep[e.g.][]{massey89a,massey89b,hill1994,massey95,hill95,massey98}
one can see that the corresponding IMF slopes in star-forming regions of
the LMC are found clustered around the value $x\approx 1.5$, similar to
the Galactic average IMF. The system IMF we derive here for LH~95 for stars
within the same mass limits does not differ significantly from this
value. Extensive studies have also shown that the high-mass IMF is
independent of metallicity in the Galaxy and the Magellanic Clouds
\citep[see review by][]{massey03}, and our results also suggest a weak
metallicity dependence of the high-mass IMF, providing further support
to its universality, in the sense that the {\sl average} IMF of a galaxy
seems to be independent from environmental factors, and that any
individual stellar system is not expected to share the exact IMF.
Indeed, the standard galactic IMF, as presented by, e.g., \citet{scalo}
and \citet{kroupascience}, is the average over a set of systems, each of
them showing slightly different mass distributions.

\begin{figure}
\epsscale{1.15}
\plotone{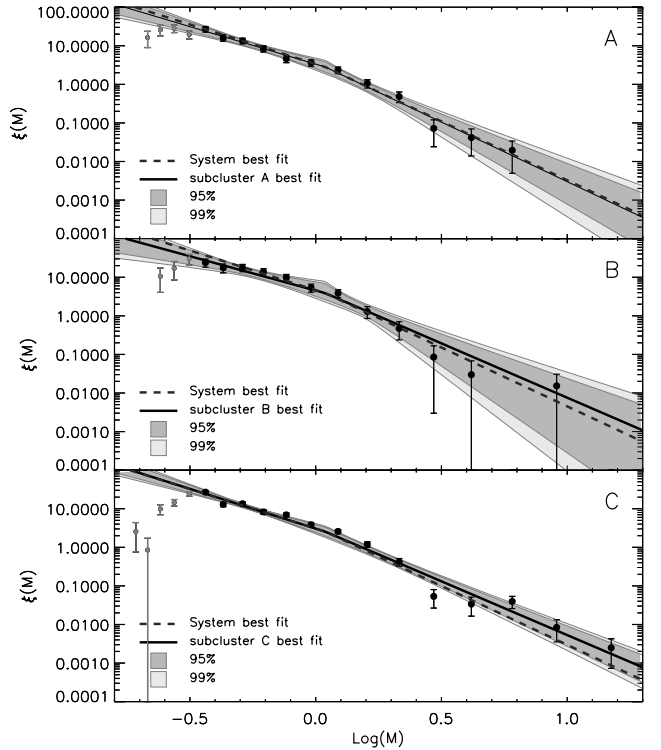}
\caption{Measured mass function for the 3 subclusters of LH~95, as they
are defined in Table \ref{t:subcluster} and shown in
Figure~\ref{fig:densitymap}b. The method used for the determination of the
best fitted two-phase power-law is analogous to that applied for the
entire central region of LH~95, except for the fact that here we constrain
the break point to be the same derived for the global IMF, at
$\log(M)=0.04$. The overall $\xi(M)$ of LH~95 is overlayed as a dashed
line.\label{fig:IMF_subclusters}}
\end{figure}

\subsection{Variability of the IMF within LH~95}

The clustering behavior of the PMS stars in LH~95 is discussed in
\S~\ref{section:centralregion}, where three high concentrations of such
stars within the main body of the system are identified as the
sub-clusters of the association LH~95 (given the names cluster A, B and
C respectively). In order to study how the IMF may vary from one
sub-cluster to the other, and how the IMF of each individual
concentration of PMS compares to the ``overall'' IMF of the system, we
constructed the IMF for each of the three sub-clusters.

We applied the same method used in \S\ref{section:IMFconstruction} and
\ref{section:IMFfitting}, isolating each time the regions of the
sub-clusters, as shown in Figure~\ref{fig:densitymap}b, iterating the
Monte Carlo field subtraction, computing the distribution of number of
stars in mass bins and correcting for incompleteness. For the model
fitting, we used the same Levenberg-Marquard technique, with the
difference that, in this case, we constrained the position of the break
point between the two power-laws to be the same as for the general IMF,
namely, at $\log(M/M_\odot)=0.04$. This choice is reasonable considering
that we are mainly interested in eventual changes in the derived slopes
of the IMF.

The constructed IMF for each sub-cluster is shown in
Figure~\ref{fig:IMF_subclusters}. The overall IMF for the whole system is
also overlayed, after it is normalized in order to match the
subclusters' best fit on the break point. In each case the IMF for the
subsolar masses is the same as the general IMF, as
discussed in \S\ref{section:IMFdiscussion} for the entire system of
LH~95. Moreover, the derived two-phase models that fit the IMF of each
sub-cluster and the corresponding uncertainties, given in
Table~\ref{t:IMF_expression_subcluster}, are compatible to that of the
overall IMF given in Table \ref{t:IMF_expression}.

\begin{deluxetable}{rcc}
\tablewidth{0pt}
\tablecaption{IMF slopes and uncertainties for the three sub-clusters
of LH~95 \label{t:IMF_expression_subcluster}}
\tablehead{
\colhead{}&
\colhead{$-0.5<\log(M/M_\odot)<0.04$}&
\colhead{$\log(M/M_\odot)>0.04$} \\
}
\startdata
subcluster A & $0.94^{-0.40}_{+0.41}$ & $2.09^{-0.55}_{+0.70}$ \\
\\
subcluster B & $0.78^{-0.77}_{+0.53}$ & $1.82^{-0.60}_{+1.71}$ \\
\\
subcluster C & $1.05^{-0.35}_{+0.33}$ & $1.79^{-0.29}_{+0.36}$ \\
\enddata
\tablecomments{slopes are in terms of $x$ according to Equation \ref{equation:imfdef}, uncertainties are $95\%$ confidence intervals.}
 \end{deluxetable}

The IMF constructed for each sub-cluster in LH~95 suggests that the mass
distribution of the PMS stars in this association does not vary much with location. Moreover, the spatial distribution of the
intermediate- and high-mass stars within the whole of the system does
not show any indication that the system is mass segregated. On the other
hand if the individual sub-clusters are considered, the more massive
stars do occupy the central parts of each of them, providing clear
evidence that mass segregation in LH~95 probably occurs in rather small
clustering scales.

According to recent results by \citet{weidner06} on the integrated
galaxial stellar IMF, the overall IMF of a composite system should be
steeper than the one it would have if it was not composite, in the
intermediate- and high-mass regime. Taking into account the fact that
LH~95 is resolved into at least three individual subclusters
(\S\ref{section:centralregion}), it may be considered as a scaled-down
case of the scenario of these authors, and therefore as a composite
system. It should be noted that also this may explain the somewhat steep
slope of the IMF we derive for LH~95.

\section{Conclusions}\label{section:conclusions}

We present a comprehensive investigation of the stellar populations of
the star-forming association LH~95, located on the north-western edge of
the super-giant shell LMC~4 in the Large Magellanic Cloud, using of
deep HST/ACS photometry in the filters $F555W$ and $F814W$ (proxies for
standard $V$ and $I$ respectively). Our findings can be summarized as
follows.

We construct a new set of observational PMS models (both isochrones and
evolutionary tracks) by starting from the original computations by
\citet{siess2000} and performing synthetic photometry on artificial
stellar spectra. The latter were derived from interpolation of
atmosphere models from the \textsc{NextGen} \citep{nextgen} and Kurucz
\citep{kurucz93} grids. This approach enabled several improvements
considering observed data for PMS stars. Specifically:

\begin{itemize}

\item[(i)] We computed absolute magnitudes for each point of the
original theoretical models in 88 different photometric bands,
including the ACS/WFC system, which is useful for our observations. The
availability of observational PMS evolutionary models in any given
photometric system allows us to avoid the application of any color
corrections to the observed data, or to approximate observations in the
ACS/WFC system with any standard photometric system.

\item[(ii)] Our computations, being similar to that of \citet{girardi02}, do
not rely on a single dependence on effective temperature
for the computation of colors and magnitudes, but they also take into account the dependencies on
the stellar surface gravities (i.e. in luminosity), and therefore they
provide a more accurate treatment of the young stages of PMS evolution.

\item[(iii)] These observational models are extrapolated to lower
metallicities than those of \citet{siess2000} by computing our synthetic
photometry for the average LMC and SMC metallicities of
$Z=0.008$ and $Z=0.004$, respectively.

\end{itemize}

We enhanced the catalog of stars detected in the observed field around
the association LH~95 with ACS \citep{lh95first} by identifying new point sources, which were previously rejected by the photometry
mostly in crowded regions. Therefore, about $900$ additional stars are
included in our photometric catalog (Table 1), which now includes in
total 17,245 stars. Our deep ACS photometry allowed us to analyze the
properties of the abundant PMS population found in LH~95, which shows an average age of 4 Myr. We performed
star counts on our stellar sample of the area of the association and we
selected the central part of the observed field, where a statistically
significant concentration of PMS stars is found, as the most
representative region of LH~95. The 3$\sigma$ isopleth was selected to
define the boundaries of the system. The PMS stars within the
association are found to be mostly concentrated in three distinctive
sub-clusters, named cluster A, B and C respective, providing evidence of
the existence of stellar subgroups within a single association.

We estimate the total stellar
masses and corresponding dynamical timescales for the system itself as
well as the three subclusters, and we find that the system cannot be
considered dynamically evolved, and it does not experience any disruption.
Moreover, the fact that LH~95 is a bound cluster with its gas partially
removed implies that the star formation efficiency is probably higher than
the typical values for young clusters.

We estimated the stellar contribution of the general background field of
the galaxy in the area of the system from our ACS observations of a
nearby control field. We then performed field subtraction from the
area of the system with the use of Monte Carlo simulations.

With our evolutionary models we estimated the interstellar
extinction in LH~95 and we computed the masses of all the members of the system. Our stellar sample reaches masses as low as
$\sim$~0.2~M{\solar}, and we are able to correct the incompleteness
accurately down to $\simeq$~0.43~M$_\odot$. For the first time, this allows the construction of the extragalactic IMF down to the sub-solar mass
regime.

We derive the system IMF by counting stars within bins of variable width, which
increase with mass, and correcting for incompleteness based on
artificial star experiments for the stars included in each mass range.
We derive the actual statistical distribution of the counting errors
associated to each mass bin via a Monte Carlo technique, in order to
take into account uncertainties introduced by the contamination from the
field population. The statistical {\sl F test} for model selection
allowed us to determine a two-phase power-law as the most appropriate model to fit our IMF. We detect a change in the slope of the IMF (the
``knee'' of the IMF) for masses lower than 1~M$_\odot$. A
Levenberg-Marquard algorithm for least square fitting was applied to
determine the power-law exponents and the point where the slope changes.
The derived system IMF slope is found to be $x=1.05^{-0.20}_{+0.15}$ in the
subsolar regime, and $x=2.05^{-0.28}_{+0.39}$ at intermediate and high
masses, with a $95\%$ of confidence.

This {\sl system} IMF, once corrected for unresolved binarity, is statistically compatible, within the studied mass
range, with the average Galactic IMF, concerning its slope and the mass limit where it changes (1~M{\solar}).
As a consequence, we cannot consider the metallicity difference between the LMC ($Z=0.008$) and the Galaxy important enough for a significant change in the IMF slope.
Certainly similar data on such
stellar systems will show if this is a normal behavior of the IMF in
this galaxy. We find no significant differences in the shape of the
overall IMF of LH~95 from that of each of the three individual PMS
sub-clusters of the association. This clearly suggests that the IMF of
LH~95 is not subject to local variability.

\acknowledgements

N.~Da Rio kindly acknowledges financial support from the German
Aerospace Center (DLR) through Grant 50~OR~0401. D. A. Gouliermis
acknowledges the support of the German Research Foundation (DFG) through
Grant 1659/1-1.

\end{document}